\documentclass[conference]{IEEEtran}
\ifCLASSINFOpdf
\else
\fi
\usepackage{booktabs} 
\usepackage{algorithm}
\usepackage{algorithmic}
\usepackage{diagbox}
\usepackage{caption} 
\usepackage{amsmath,amsfonts,amssymb,amsthm}
\usepackage{bm}
\usepackage{graphicx}
\usepackage{subfigure}
\usepackage{capt-of}
\usepackage{multirow,array}
\usepackage{color}
\usepackage{tikz}
\usepackage{pbox}
\usepackage{url}

\newcommand*\circled[1]{\tikz[baseline=(char.base)]{
            \node[shape=circle,draw,inner sep=0.5pt] (char) {\tiny #1};}}

\newcommand{\bx}{\bm{x}}

\newcommand{\bb}{\bm{b}}

\newcommand{\bo}{\bm{o}}
\newcommand{\bz}{\bm{z}}

\newcommand{\bW}{\bm{W}}

\newcommand{\bth}{\bm{\theta}}

\begin{document}
%
\title{A Deep Learning-based Framework for Conducting Stealthy Attacks in Industrial Control Systems}



%
\author{\IEEEauthorblockN{Cheng Feng\IEEEauthorrefmark{1},
Tingting Li\IEEEauthorrefmark{1},
Zhanxing Zhu\IEEEauthorrefmark{2} and
Deeph Chana\IEEEauthorrefmark{1}}
\IEEEauthorblockA{\IEEEauthorrefmark{1}Institute for Security Science and Technology\\
Imperial College London,
London, United Kingdom \\ Emails: \{c.feng, tingting.li, d.chana\}@imperial.ac.uk}
\IEEEauthorblockA{\IEEEauthorrefmark{2}Peking University and BIBDR, Beijing, China\\
Email: zhanxing.zhu@pku.edu.cn}}



\maketitle

\begin{abstract}
Industrial control systems (ICS), which in many cases are components of critical national infrastructure, are increasingly being connected to other networks and the wider internet motivated by factors such as enhanced operational functionality and improved efficiency. However, set in this context, it is easy to see that the cyber attack surface of these systems is expanding, making it more important than ever that innovative solutions for securing ICS be developed and that the limitations of these solutions are well understood. The development of anomaly based intrusion detection techniques has provided capability for protecting ICS from the serious physical damage that cyber breaches are capable of delivering to them by monitoring sensor and control signals for abnormal activity. Recently, the use of so-called \emph{stealthy attacks} has been demonstrated where the injection of false sensor measurements can be used to mimic normal control system signals, thereby defeating anomaly detectors whilst still delivering attack objectives. To date such attacks are considered to be extremely challenging to achieve and, as a result, they have received limited attention.

In this paper we define a deep learning-based framework which allows an attacker to conduct stealthy attacks with minimal \emph{a-priori} knowledge of the target ICS. Specifically, we show that by intercepting the sensor and/or control signals in an ICS for a period of time, a malicious program is able to automatically learn to generate high-quality stealthy attacks which can achieve specific attack goals whilst bypassing a black box anomaly detector. Furthermore, we demonstrate the effectiveness of our framework for conducting stealthy attacks using two real-world ICS case studies. We contend that our results motivate greater attention on this area by the security community as we demonstrate that currently assumed barriers for the successful execution of such attacks are relaxed. Such attention is likely to spur the development of innovative security measures by providing an understanding of the limitations of current detection implementations and will inform \emph{security by design} considerations for those planning future ICS. 
\end{abstract}


%

\section{Introduction}

Industrial Control Systems (ICS) generally consist of a set of
supervisory control and data acquisition (SCADA) subsystems for
controlling field devices via the monitoring and processing of data
related to industrial processes. In response to information received
from remote sensors, control commands are issued either automatically,
or manually, to remotely located control devices, which are able to
make physical changes to the states of one or more industrial
processes. In the pursuit of increased communication efficiency and
higher throughputs, modern information and communication technologies (ICT) have been widely integrated
into ICS. For instance, cloud computing has recently emerged as a new
computing paradigm for ICS, proposing the conversion of localised data
historians and control units into cloud based services
\cite{givehchi2013cloud}. Such an evolution satisfies the growing
needs from operators, suppliers and third-party partners for remote
data access and command execution from multiple platforms. A trend
which,  consequently, exposes modern ICS to an increased risk of
cyber attacks. Unlike conventional ICT systems, ICS security breaches
have the potential to result in significant physical damage to
national infrastructure; resulting in impacts that can include large
scale power outages, disruption to health-service operations,
compromised public transport safety, damage to the environment and direct
loss of life.

The first well known ICS-targeted cyber virus  -- \emph{Stuxnet}
\cite{stuxnet} -- popularised cyber security vulnerabilities of ICS
and demonstrated how external attackers might feasibly penetrate
multiple layers of security to manipulate the computer programs that
control their field devices. Since then, according to sources such as
ICS-CERT, the number of attacks recorded against ICS targets has grown
steadily.  In 2014 \cite{icscert2014} 245 incidents were reported to
ICS-CERT by trusted industrial partners. This figure increased to 295
in 2015 \cite{icscert2015} and then again to 290 by 2016
\cite{icscert2016}. More recently than Stuxnet, a major security
breach to a German steel mill was widely reported in 2014. This attack
was initiated by credential theft via spear-phishing emails, leading
to massive damage to a blast furnace \cite{germansteelmill} at the
plant and illustrating clearly the cyber-physical nature of ICS
security. Even more recently, in December 2015, Ukrainian's capital
city Kiev reportedly lost approximately one-fifth of its required
power capacity as a result of a utility-targeted cyber attack that
caused a massive blackout, affecting 225,000 citizens \cite{ukrainian}.

%

As an approach to implementing effective intrusion detection systems
(IDS), anomaly detection has become recognised as part of the standard
toolkit for protecting ICS from cyber attacks. In recent years many
ICS-specific anomaly detection models have been proposed \cite{mitchell2014survey,zhu2010scada}, mainly
incorporating blacklist-based or whitelist-based methods. These IDS
are designed to detect intrusions by consideration of common ICS
communication protocols (e.g. Modbus/DNP3) \cite{fovino2010modbus}, ICS operating standards \cite{kang2016csr,yang2013intrusion}
and the identification of patterns and structures of data transmitted
across ICS networks. Amongst the various underpinning technologies used
in existing IDS, machine learning (ML) techniques have enabled
detecting modalities that exploit automated learning from past
experience and the construction of self-evolving models that can adapt
to evolving/future classification problems. To date, ML approaches
that have shown promising capabilities for securing ICS include: Support Vector Data Description (SVDD)
\cite{nader2014norms}, Bloom filters \cite{parthasarathy2012bloom},
statistical Bayesian networks \cite{bigham2003safeguarding} and deep neural networks
\cite{goh2017anomaly,feng2017multi}, etc.


Furthermore, anomaly detection techniques employed at the field
network layer provide protection at a point within ICS where security
compromise has the greatest potential for causing significant physical damage
to the system. At present, most
anomaly detection implementations designed for this specific purpose
rely on a predictive model, where future sensor measurements
predictions are generated based on historical signals and compared
with real measurements using a thresholded residual error. Comparisons
exceeding the defined threshold constitute a detection event and
generate an
alert \cite{van1985bad,abur2004power,hadvziosmanovic2014through,goh2017anomaly}. However,
recent studies have shown that cyber attacks may be generated by
the injection of false sensor measurements that avoid the detection by such systems
\cite{dan2010stealth,liu2011false}. Needless to say, such
\emph{stealthy} attacks are amongst the most harmful class of attacks
against ICS that exist, as malicious sensor measurements can directly violate
the operational safety bounds and conditions set for a given
ICS. Although efforts to understand the risks
and mitigations related to stealthy attacks on ICS have been made
\cite{teixeira2012revealing,urbina2016limiting}, it is still
generally believed that this class of attacks has a low likelihood
of occurrence. This is due, in most part, to the level and type of
operational information of the system that is thought to be needed by
an attacker \emph{a-priori};
information that is challenging to obtain.

In this work we challenge the above \emph{strong} conditions imposed on a
stealthy attacker by formulating and demonstrating an attack framework that uses a deep learning-based methodology. The deep learning method proposed allows the attacker to effectively conduct stealthy attacks which lead to \emph{a specific amount} of deviation on sensor measurements with minimal knowledge of the target ICS. Specifically, our framework
consists of set of deep neural network models, which are trained by a recently
proposed adversarial training technique -- Wasserstein GAN
\cite{arjovsky2017wasserstein}, a special variant of Generative Adversarial Net (GAN) \cite{goodfellow2014generative} to maximize
the likelihood that generated malicious sensor measurements will
successfully bypass an assumed black-box anomaly detector. This
technique significantly lowers the bar for conducting stealthy
attacks as generating high-quality attacks can be
automatically achieved by intercepting the sensor and/or control
signals in an ICS for a period of time using a particularly designed
real-time learning method. In addition, the explicit relaxation of
conditions relating to specific operational knowledge within the
attack design demonstrates the potential for its general applicability across ICS
instances and industrial sectors. To show this we demonstrate the effectiveness of our
framework by two case studies in which we conduct stealthy attacks on
a gas pipeline and a water treatment system, respectively. Overall, our results
indicate that  more attention to stealthy attacks is urgently required
within the ICS community.


\section{Background}
By way of providing relevant background for the work undertaken here,
this section provides an brief overview of the ICS control network
in Section \ref{sec:ids}. The generation and detection of stealthy attacks is briefly discussed in Section \ref{sec:gen}. Following this, Section \ref{sec:deep} provides
information on the key underpinning techniques of deep neural networks
employed here.

\subsection{ICS Control Network}\label{sec:ids}

\begin{figure}[t]
  \centering
  \includegraphics[width=0.99\columnwidth]{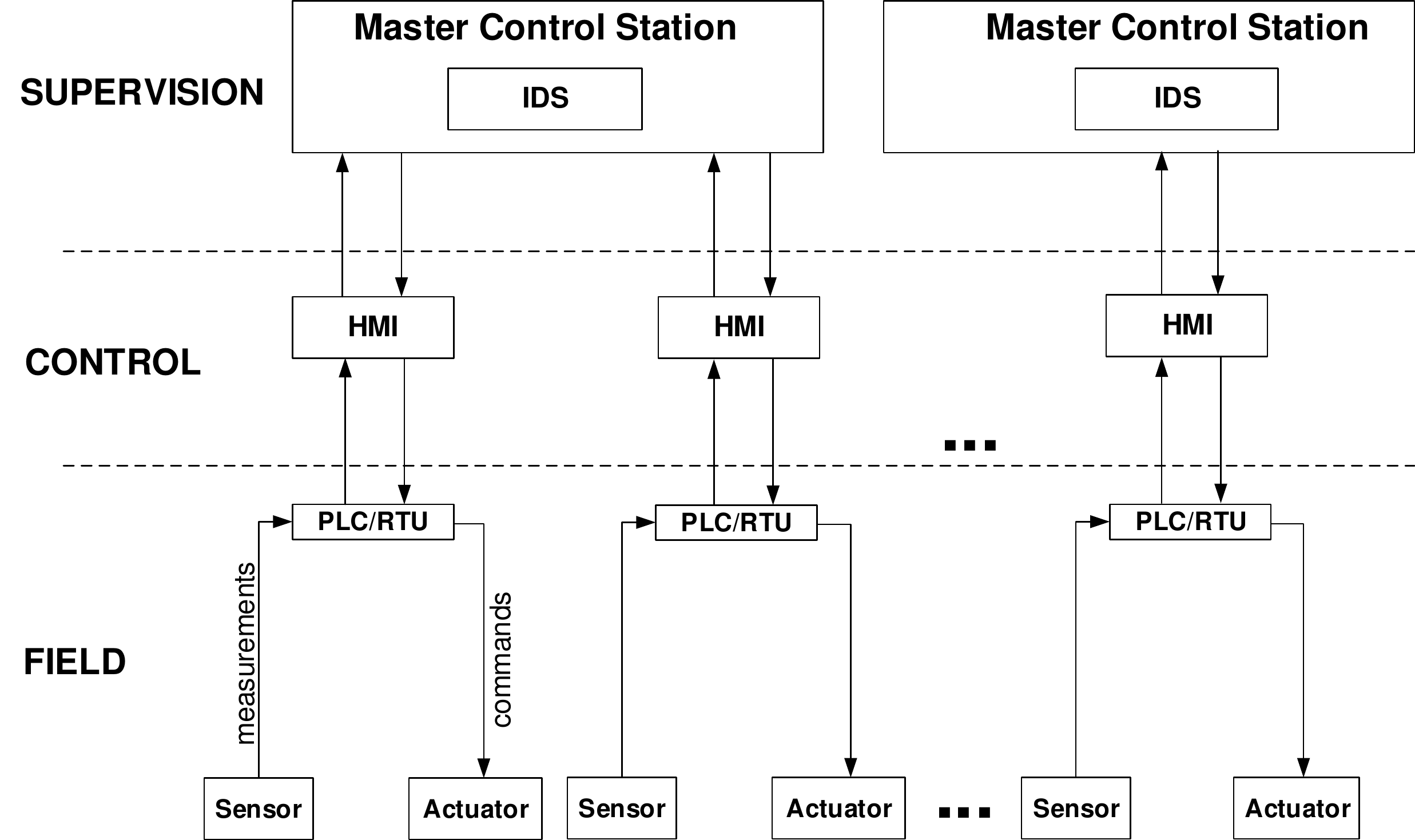}\\
  \caption{ICS control network}\label{fig:intro}
\end{figure}

We summarize the main control loops of a typical ICS in Figure \ref{fig:intro}, showing the main communication transactions within it. Specifically, in the field network level, programmable logic controllers (PLCs) or Remote Terminal Units (RTUs) receive real-time measurements from sensors and forward them to the Human Machine Interface (HMI) in the control network level. HMI processed data is then again forwarded to the
Master Control Station in the supervision level. In response to the received
inputs, the Master Control Station then issues necessary control commands to change the states of
its monitored industrial processes. Commands eventually make their way
to relevant actuators to make the physical changes required. Intrusion detection systems (IDS) are often deployed on the master control stations, where the sensor measurements and control signals are monitored to secure the physical processes under control. We refer to Section~\ref{sec:mechanism} for the detailed description about the mechanisms for securing physical processes in ICS.

\subsection{Generating and Revealing Stealthy Attacks} \label{sec:gen}

Stealthy attacks, or more generally mimicry attacks, have been studied in the
community of conventional IT security for decades. Such attacks can achieve specific attack goals without
introducing illegal control flows into computer programs. Existing IDS
for computer programs were mainly relied on short call sequence verification,
which were unable to detect such attacks. The work in
\cite{shu2015unearthing} proposed an IDS implemented by two-stage
machine learning algorithms to construct normal call-correlation
patterns and detect stealthy intrusions. In the domain of monitoring
the behaviours of applications,  system calls have been used
extensively to construct the normal profiles for intrusion detection
\cite{kruegel2005automating,wagner2002mimicry}. However, it has
been shown that such IDS are inadequate to detect mimicry attacks
\cite{wagner2002mimicry} (i.e. attacks that interleave the malicious
code with legit code) and impossible paths attacks
\cite{xu2004context} (i.e. attacks relying on a legal yet never
executed sequence of system calls). The work presented in
\cite{wagner2002mimicry} was one of the earliest studies on mimicry
attacks against host-based IDS, where a theoretical framework was
proposed to evaluate the effectiveness of an IDS combating mimicry
attacks. A waypoint-based IDS was introduced in \cite{xu2004context}
to detect both mimicry attacks and impossible paths attacks by
considering the trustworthy execution contexts of programs and
restricting system call permissions. Besides, there were also existing
work conducted on generating mimicry attacks by using generic
programming \cite{kayacik2009generating} or static binary analysis
\cite{kruegel2005automating}.

In many successful ICS attacks, physical damage is achieved from an
exploitation phase that utilises methods of injecting \emph{false}
sensor measurements into the control network. As discussed briefly previously, such attacks
have also been described as \emph{Stealthy Attacks}. An early practical description of such an attack is
provided in \cite{liu2011false}. Here the ability of an attacker
to insert arbitrary errors into the normal operation of a power system
without being detected by an implemented state estimator IDS is
shown. It was pointed out that launching such attacks do pose strong requirements for the attackers as the configuration of the targeted power system must be known. Further aspects of stealthy attacks have been studied in
\cite{dan2010stealth,sandberg2010security}. Specifically, two
security indices were introduced in \cite{sandberg2010security} to quantify the
difficulty of launching successful stealthy attacks against
particular targets and the work in
\cite{dan2010stealth} has proposed an efficient way to compute such
security indices for sparse stealthy attacks. A stealthy deception
attack against a canal system is shown in \cite{amin2013cyber} and the
effect of stealthy attacks on both the regulatory layer and the
supervisory layer of an ICS is discussed. In \cite{teixeira2012revealing}, the authors proposed to reveal stealthy attacks in control systems through modifying the system's structure periodically. Recently, the authors in \cite{urbina2016limiting} showed that the impact of stealthy attacks can be mitigated by the proper combination and configuration of different off-the-shelf detection schemes.

\subsection{Deep Neural Networks}\label{sec:deep}
Recently, deep learning~\cite{lecun2015deep} has achieved remarkable
success and improved the state-of-the-art in various artificial
intelligence applications including visual object classification,
speech recognition and natural language understanding. It is a
powerful modeling framework in machine learning, where multiple
processing layers are used to extract features from data with multiple
levels of abstraction. In this work, several neural
network architectures are used to conduct feature extraction, anomaly detection
and stealthy attacks generation.
We describe these architectures below.
\subsubsection{Feedforward Neural Network (FNN)}
\begin{figure}
 \centering
 \includegraphics[width=0.99\columnwidth]{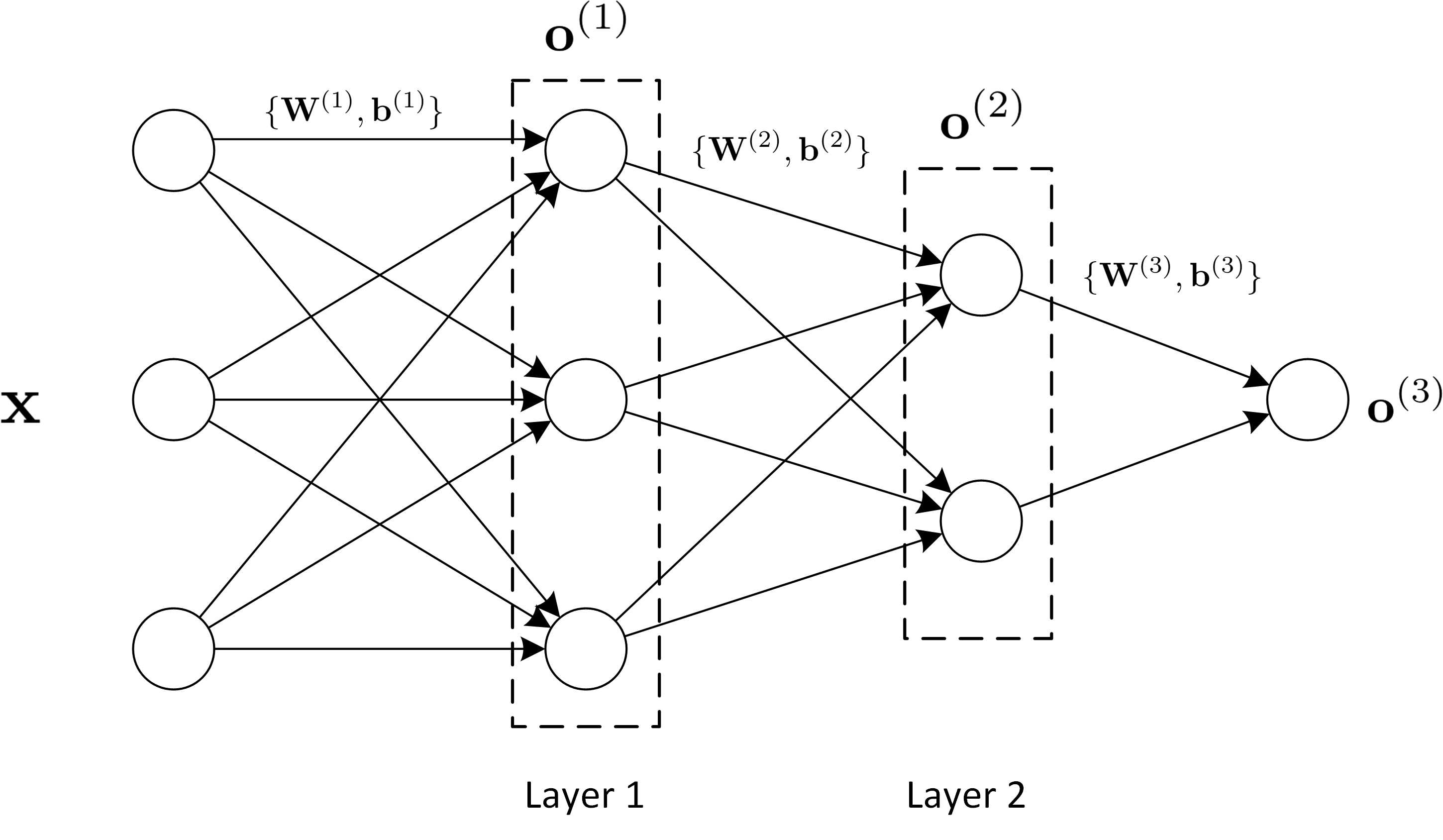}
 \caption{The architect of a typical FNN}
 \label{fig:fnn}
\end{figure}

The FNN, also called multi-layer perceptron
(MLP), describes the most classic form of neural network (NN) where
multiple processing nodes are arranged in layers such that information
only flows in one direction -- from input to output. The architect of a typical FNN is illustrated in Figure~\ref{fig:fnn}.  In this
architecture the $j$-th node in $l$-th layer computes linear
combination of its inputs
(i.e. the outputs of last layer) followed by a simple non-linear
transformation $f(\cdot)$,
\begin{equation*}
 o_j^{(l)} = f\left(\sum_{i} W^{(l)}_{ij} o_i^{(l-1)} + b^{(l)}_j \right),
\end{equation*}
where we often use rectified linear unit (ReLU) as the non-linear
function $f(z) = \max (0,z)$, the input $\bx = \bo^{(0)}$. And the
model parameters $\bth = \{ \bW^{(l)}, \bb^{(l)} \}_{l=1}^L$ needs to
be learned by minimizing certain loss function
\begin{equation*}
 J(\bth) = \frac{1}{N}\sum_{i=1}^N e (o^{(L)}(\bx_i), y_i; \bth) ,
\end{equation*}
given the training data $\{(\bx_i, y_i)  \}_{i=1}^N$, where
$e(\cdot,\cdot)$ is some chosen criterion to measure the difference
between the prediction and ground truth, such as squared loss for
regression and cross entropy for classification problem. Often
stochastic gradient descent (SGD) can be used to minimize the loss $J(\bth)$. Concretely, in each iteration, a mini-batch of training samples are selected randomly to estimate the true gradient, $\nabla_{\bth} \tilde{J} = \sum_{j=1}^m \nabla_{\bth} e (o^{(L)}(\bx_j), y_j; \bth)$, where $m$ is the number of samples inside each mini-batch. Then we can update the parameters by $\bth \leftarrow \bth - \alpha \nabla_{\bth} \tilde{J}$ with some given learning rate $\alpha$.

\subsubsection{Recurrent Neural Network (RNN)}
\begin{figure}[t]
 \centering
 \includegraphics[width=0.99\columnwidth]{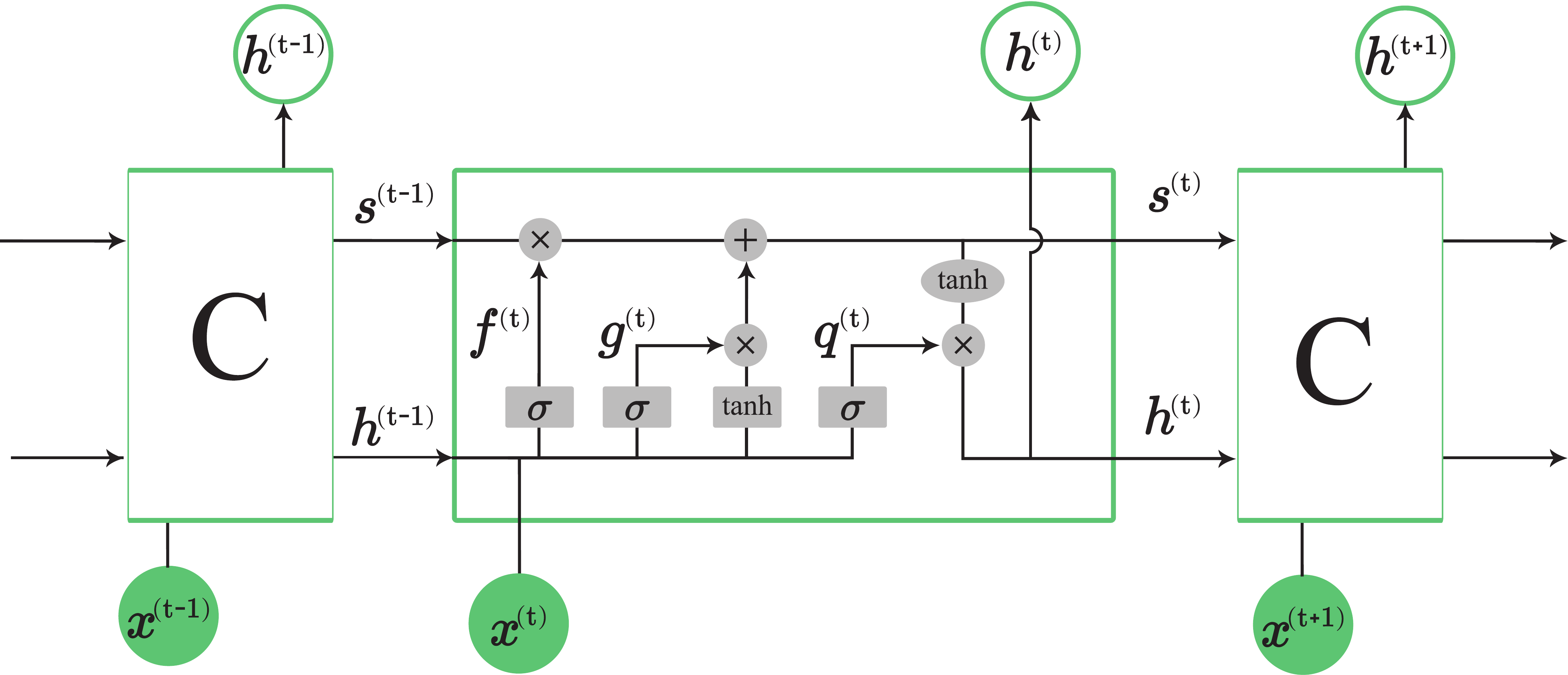}
 \caption{LSTM cell}
 \label{fig:lstmcell}
\end{figure}

In contrast to FNNs, RNNs permit cyclical connections between nodes
allowing such neural networks to exhibit dynamic temporal properties
in operation. RNNs~\cite{lipton2015critical} are suitable for dealing with
sequential data, such as speech, language and structured time series.
An input sequence is processed by an RNN one element at a time, and the
information is encoded into a hidden unit (i.e. a state vector) for
each time step that describes the history of all the past elements of
the sequence. The outputs of the hidden units at
different time steps can be compared with the ground truth of the
sequence such that the network can be trained. The most commonly
implemented RNNs fall into the class of long short-time memory (LSTM~\cite{hochreiter1997}) neural
networks. As the name suggests, such NNs exhibit remarkable empirical
performance for extracting/preserving long-term dependencies whilst
also maintaining short-term signals. LSTM networks involve three gates
in the computation of each hidden cell to determine what to forget,
what to output and what to be provided to next hidden cell,
respectively, as shown in Figure~\ref{fig:lstmcell}. The information flow of LSTM cell
is as follows,
\begin{small}
\begin{align}
 \bm{f}^{(t)} &= \sigma \Big( \bm{b}^f + \bm{U}^f \bm{x}^{(t)} +  \bm{W}^f \bm{h}^{(t-1)} \Big), \label{eq:dl:fg}\\
 \bm{g}^{(t)} &= \sigma \Big( \bm{b}^g +  \bm{U}^g \bm{x}^{(t)} +  \bm{W}^g \bm{h}^{(t-1)} \Big), \label{eq:dl:eig}\\
 \bm{q}^{(t)} &= \sigma \Big( \bm{b}^{q} +  \bm{U}^q \bm{x}^{(t)} + \bm{W}^q \bm{h}^{(t-1)} \Big), \label{eq:dl:og}\\
  \bm{s}^{(t)} &= \bm{f}^{(t)} \odot \bm{s}^{(t-1)} +  \bm{g}^{(t)} \odot
 \tanh \Big( \bm{b} + \sum_j \bm{U} \bm{x}^{(t)} +  \bm{W} \bm{h}^{(t-1)} \Big), \label{eq:dl:lstms}\\
 \bm{h}^{(t)} &= \tanh\big( \bm{s}^{(t)} \big) \odot \bm{q}^{(t)}, \label{eq:dl:lstmh}
\end{align}
\end{small}
where  $\sigma(\cdot)$ and $\tanh(\cdot)$ represent the sigmoid and hyperbolic tangent function, respectively,  and $\odot$ denotes the element-wise product.

\subsubsection{Generative Adversarial Net (GAN)} GANs~\cite{goodfellow2014generative} are an example of generative models that aim to learn an estimate (i.e. $p_{\text{model}}(\bx)$) of the data distribution $p_{\text{data}}(\bx)$, given the training samples drawn from  $p_{\text{data}}(\bx)$, such that we can generate new samples from the $p_{\text{model}}(\bx)$.

The general idea of GANs is to construct a game between two players. One of them is
called the generator, which intends to create samples following the same distribution as the training data. The other player is the discriminator which tries to distinguish whether the samples (obtained from the generator) are real or fake. When the discriminator cannot tell apart generated and training samples, implying that we have learned the data distribution, i.e. $p_{\text{model}}(\bx) = p_{\text{data}}(\bx)$.

The generator is simply a differentiable function $\bx = G(\bz; \bth_{G})$ with model parameters $\bth_{G}$, where the input $\bz$ follows a simple prior distribution, such as uniform and Gaussian distribution. Due to the high capacity of  multi-layer neural networks, they  are often used as the $G(\cdot)$ function, and its output $\bx \sim p_{\text{model}}(\bx)$. The discriminator is a two-class classifier, $D(\bx; \bth_{D})$ often designed as a neural network to output class probability between 0 and 1. GAN aims to solve the following min-max optimization problem,
\begin{small}
\begin{equation*}
 \min_{\bth_G} \max_{\bth_D} \mathbb{E}_{\bx \sim p_{\text{data}}(\bx)}[\log D(\bx;\bth_D)] + \mathbb{E}_{\bz \sim p_{\bz}(\bz)}[\log(1-D(G(\bz;\bth_G)))]
\end{equation*}
\end{small}
An alternative update of $\bth_D$ and $\bth_G$ by SGD can be adopted to solve this problem. Then we can use the optimized $\bx = G(\bz; \bth_{G}^*)$ to generate new samples through the random input $\bz$.

Unfortunately, the training of the original GAN is very instable in practice, and users have to balance both of the capacity and training steps between the generator and discriminator. To overcome this issue, the authors of~\cite{arjovsky2017wasserstein} proposed to apply Wasserstein distance to measure the discrepancy between the two probability distributions, hence the name, Wasserstein GAN (WGAN).  The min-max optimization problem of WGAN can be formulated as,
\begin{small}
\begin{equation}
 \min_{\bth_G} \max_{| \bth_D|_{\infty} \leq c} \mathbb{E}_{\bx \sim p_{\text{data}}(\bx)}[ D(\bx;\bth_D)] - \mathbb{E}_{\bz \sim p_{\bz}(\bz)}[D(G(\bz;\bth_G))]. \label{eq:wgan}
\end{equation}
\end{small}
Compared with original GAN,  the discriminator $D(\cdot)$ of WGAN, also called the ``critic'', can output any real value, not just probabilities. The infinity norm of $\bth_D $ is constrained to be less than a predefined positive constant $c$, and practically this can be achieved by ``clipping'' the elements larger than $c$ to $c$ (and elements smaller than $-c$ to $-c$) in each iteration.  And the expectation term in Eq.~(\ref{eq:wgan}) can be approximated by using $m$ random samples from training data and $p_{\bz}(\bz)$.

We summarize the training procedure of WGAN in Alg.~\ref{alg:wgan_training}, where a variant of stochastic gradient descent, RMSProp~\cite{rmsprop}, is used for updating the parameters.
\begin{algorithm}
  \caption{Training Procedure of WGAN~\cite{arjovsky2017wasserstein}}
  \begin{algorithmic}[1]
  	\REQUIRE the learning rate $\alpha=0.00005$, the clipping parameter $c=0.01$, the number of iterations of the critic per generator iteration $n=5$, the size of mini-batch, $m$.
  	\REQUIRE the initial parameters $\bth_D^0$ and $\bth_G^0$.
     \WHILE{$\bth$ has not converged}
      \FOR{$t = 0,\dots, n$}
     	\STATE{Sample $\{ \bx^{(i)}\}_{i=1}^m$ a mini-batch from training data. }
     	\STATE{Sample $\{ \bz^{(i)}\}_{i=1}^m \sim p_{\bz}(\bz)$ a mini-batch of prior samples. }
     	\STATE{Evaluate stochastic gradient of $\bth_{D}$:
     	\begin{scriptsize}
      	 \begin{eqnarray*}
     	\delta_{\bth_{D}} \leftarrow \nabla_{\bth_{D}}  [  \sum_{i=1}^m D(\bx^{(i)}; \bth_{D})/m - \sum_{i=1}^m D(G(\bz^{(i)};\bth_{G}); \bth_{D})/m  ]
     	\end{eqnarray*} \end{scriptsize}}
     	\STATE{$\bth_D \leftarrow \bth_D + \alpha \cdot RMSProp(\bth_D, \delta_{\bth_D})$}
     	\STATE{$\bth_D \leftarrow clip(\bth_D , -c, c)$}
     \ENDFOR
     \STATE{Fix $\bth_D$ }
     \STATE{Sample $\{ \bz^{(i)}\}_{i=1}^m \sim p_{\bz}(\bz)$ a mini-batch of prior samples. }
     \STATE{Evaluate stochastic gradient of $\bth_{G}$:
     $$ \delta_{\bth_G} \leftarrow -\nabla_{\bth_G}  \sum_{i=1}^m D(G(\bz^{(i)};\bth_{G}); \bth_{D})/m $$}
      \STATE{$\bth_G \leftarrow \bth_G + \alpha \cdot RMSProp(\bth_G, \delta_{\bth_G})$}
      \ENDWHILE
  \end{algorithmic}
   \label{alg:wgan_training}
\end{algorithm}

\section{Anomaly Detection Mechanism for Securing Physical Processes}
\label{sec:mechanism}
As discussed in Section~\ref{sec:ids}, the physical process in ICS is directly controlled by the field network which consists of a number of field devices called sensors, actuators and PLCs. Specifically, the sensors are devices which convert physical parameters into electronic measurements; actuators are devices which convert control commands into physical state changes (e.g., turning a pump on or off); based on measurements received from sensors through PLC-sensor channels, PLCs send control commands to actuators through PLC-actuator channels.

To protect the control system from physical faults and cyber attacks, anomaly detection mechanisms are often deployed by monitoring the sensor measurements and control commands in the system at discrete time steps. Concretely, let $\mathbf{X} = \{ \mathbf{x}^{(1)}, \mathbf{x}^{(2)}, \ldots \}$ be a time-series in which each signal $\mathbf{x}^{(t)}=\{ \mathbf{y}^{(t)},\mathbf{u}^{(t)} \} \in \mathbb{R}^{m+n}$ is a $(m+n)$-dimensional vector
\begin{equation*}
\{ y^{(t)}_1, y^{(t)}_2, \ldots, y^{(t)}_m, u^{(t)}_1, u^{(t)}_2, \ldots, u^{(t)}_n \}
\end{equation*} whose elements correspond to $m$ values representing all the sensor measurements and $n$ values capturing all the control commands in the system at time $t$. Currently, most anomaly detection mechanisms for the control system rely on a predictive model which predicts the sensor measurements $\hat{\mathbf{y}}^{(t)} = \{ \hat{y}^{(t)}_1, \hat{y}^{(t)}_2, \ldots, \hat{y}^{(t)}_m \}$ based on previous signals, and an alarm is triggered if the residual error between the predicted measurements and the true measurements exceeds a specific threshold.

\subsection{Predictive Models}
\label{sec:predictive}
The underlying predictive model can take many different forms, among which the Auto-Regressive (AR) model \cite{hadvziosmanovic2014through,urbina2016limiting} and the Linear Dynamic State-space (LDS) model \cite{teixeira2012revealing,urbina2016limiting} (it is often called as state estimator in power systems \cite{abur2004power}) are the most commonly used. Specifically, the AR model predicts $\hat{\mathbf{y}}^{(t)}$ by fitting a linear regression model for each sensor measurement based on its $p$ previous values:
\begin{equation*}
\hat{y}_i^{(t)}=\sum _{{j=1}}^{p}\alpha_{j}  y^{(t-j)}+\alpha_{0} \quad \forall i \in \{ 1,2,\ldots, m \}.
\end{equation*}where $\alpha_{j}, \forall j \in \{ 1, \ldots,p\}$ are coefficients representing the weights of the measurement at time $t-j$ for predicting $\hat{\mathbf{y}}^{(t)}$, $\alpha_{0}$ is a constant.

The LDS model assumes a vector $\mathbf{w} \in \mathbb{R}^k$ to denote the physical state of the system, then $\hat{\mathbf{y}}^{(t)}$ can be inferred by the following equations:
\begin{eqnarray*}
\mathbf{w}^{(t)} &=& A \: \mathbf{w}^{(t-1)} + B \: \mathbf{u}^{(t-1)} + \bm{\epsilon}^{(t-1)} \\
\hat{\mathbf{y}}^{(t)} &=& C \: \mathbf{w}^{(t)} + D \: \mathbf{u}^{(t)} + \bm{\varepsilon}^{(t)}
\end{eqnarray*}where $A,B,C,D$ are matrices capturing the dynamics of the physical system,  $\bm{\epsilon}^{(t-1)}$ and $\bm{\varepsilon}^{(t)}$ are vectors of noise for state variables and sensor measurements with a random process with zero mean. In general, $D=0$ because sensor measurements only depend on the current physical state in most systems. Then, to predict $\hat{\mathbf{y}}^{(t)}$, one can use $\mathbf{y}^{(t-1)}$ and $\mathbf{u}^{(t-1)}$ to obtain an estimate of the current system state $\hat{\mathbf{w}}^{(t)}$, and predict $\hat{\mathbf{y}}^{(t)} = C \: \hat{\mathbf{w}}^{(t)}$.

Since the system dynamics in many ICS are highly nonlinear, recently people have found that deep learning models can achieve better prediction accuracy than the linear models. For example, the authors in \cite{goh2017anomaly} shows that the LSTM model can be employed to predict:
\begin{eqnarray*}
\mathbf{h}^{(t)} &=& f(\mathbf{x}^{(t)}, \mathbf{h}^{(t-1)}) \\
\hat{\mathbf{y}}^{(t)} &=& W_y \mathbf{h}^{(t-1)} + \mathbf{b}_y
\end{eqnarray*}where $\mathbf{h}^{(t)}$ is a hidden vector computed iteratively by the first equation which encodes the previous time series signals to provide the context for predicting the sensor measurements at time $t$; $f$ is a complex function as a short form of Equations~\ref{eq:dl:fg} to \ref{eq:dl:lstmh}; $W_y$ and $\mathbf{b}_y$ represent the weight matrix and the bias vector respectively for decoding the hidden vector to the predicted sensor measurements.

\subsection{Detection Methods}
\label{sec:detect}
Based on the prediction of the predictive model, an anomalous signal can be detected when the Euclidean distance between the predicted sensor measurements and their observations exceeds a specific threshold: $ \|\hat{\mathbf{y}}^{(t)}-\mathbf{y}^{(t)}\| > \tau$, where $ \|\hat{\mathbf{y}}^{(t)}-\mathbf{y}^{(t)}\|$ is often called the residual error at time point $t$.

Instead of solely relying on the residual errors at a single time point, we could also take account of the history of residual errors. For example, we can apply the Cumulative Sum (CUSUM) method \cite{woodall1993statistical} for detecting collective anomalies. Specifically, let $r^{(t)}$ denote the residual error at time $t$, assuming residual errors at different time points are independent and identically distributed with mean $\mu$ and variance $\sigma^2$, the CUSUM method will detect anomalies based on an accumulated statistic $H^{(t)}$ such that:
\begin{eqnarray*}
H^{(t)} &=& \max(0, H^{(t-1)} + r^{(t)} - \mu - \omega ) 
\end{eqnarray*}
with initials $H^{(0)}=0$; $\omega$ is often set to a reference value such as $\sigma$. Then, if the cumulative sum $H^{(t)}$ reaches the predefined threshold $\tau$, an alarm is triggered. After an alarm is triggered, $H^{(t)}$ is set to $0$ again, and a new round of detection is initiated.

Since in many cases, anomalies are not presented in the training phase of the detection models, thus the threshold value for the residual errors is often decided by tunning the expected false alarm rate. For the CUSUM method, the expected time between false alarms can also be tuned to decide its threshold value. 

\section{Stealthy Attack Model}
In this section, we formally define the stealthy attacks which will be generated using our deep learning framework. Specifically, we consider an ICS with $l_1$ PLC-sensor channels, $l_2$ PLC-actuator channels, and an anomaly detector is monitoring the sensor measurements and control commands delivered via the channels.  Without loss of generality, we denote the anomaly detector as a function:
\begin{equation*}
\mathcal{F}(  \mathbf{y}^{(t)}\mid \mathbf{X}^{t-1}) = \begin{cases}
    1      & \quad \text{if } \mathbf{y}^{(t)} \text{ triggers an alarm }\\
    0  &  \quad \text{otherwise}\\
  \end{cases}
\end{equation*}
where $\mathbf{X}^{t-1} = \{ \ldots, \mathbf{x}^{(t-2)}, \mathbf{x}^{(t-1)} \}$ represents the monitored time series of the whole system until time $t-1$; $\mathcal{F}(  \mathbf{y}^{(t)}\mid \mathbf{X}^{t-1})=1$ indicates a bad sensor measurement is detected at time $t$.

Furthermore, we consider the attacker has the ability to intercept  $k_1$ PLC-sensor channels and $k_2$ PLC-actuator channels, where $k_1 \leq l_1$ and $k_2 \leq l_2$. Therefore, the attacker has a partial knowledge of the system dynamics, which can be denoted as a time-series $\mathbf{X}_c=\{ \mathbf{x}_c^{(1)},\mathbf{x}_c^{(2)}, \ldots  \}$ where each signal $\mathbf{x}_c^{(t)} = \{\mathbf{y}_c^{(t)}, \mathbf{u}_c^{(t)} \}$ is a subset of $\mathbf{x}^{(t)}$, consisting of the sensor measurements and control commands which are delivered via the $k_1+k_2$ compromised channels. Unlike previous works which assume the anomaly detection function $\mathcal{F}$ is known or at least partially known to the stealthy attacker \cite{dan2010stealth, liu2011false}, we assume the anomaly detector as a \emph{black box} to the attacker in this paper. 

The stealthy attacker's target is to inject malicious sensor measurements which are \emph{deviant} with their real values to \emph{a specific amount}, whilst \emph{bypassing} the black box anomaly detector $\mathcal{F}$.  Specifically, let $\tilde{\mathbf{y}}_c^{(t)}$ denote the injected malicious sensor measures at time $t$,  $\mathbf{y}_c^{(t)}$ denote their corresponding real values, we define a set of attack goals $\mathcal{G}$ for the stealthy attacker. Formally, each attack goal $g \in \mathcal{G}$ is defined as a target function:
\begin{eqnarray*}
\tilde{y}_g^{(t)} & \circled{$g$} &  v_g  \quad \text{where } \circled{$g$} \in \{ >,<,\leq,\geq,= \} \wedge \tilde{y}_g^{(t)} \in \tilde{\mathbf{y}}_c^{(t)},
\end{eqnarray*}and $v_g$ is the target compromising value set by the attacker. As an illustration, the target function $\tilde{y}_g^{(t)} < y_g^{(t)}-1$ denotes an attack goal to fool the PLC with a fake sensor measurement with more than $1$ unit smaller than its real value $y_g^{(t)}$. Clearly, such stealthy attacks are very dangerous, as they can potentially sabotage the ICS by implicitly putting them in a critical condition.




\section{Deep Learning Framework for Conducting Stealthy Attacks}
In this section, we present the methodology to automatically conduct stealthy attacks from the attackers' perspective. Specifically, the conducting of stealthy attacks consists of two phases: the reconnaissance phase and the attacking phase. In the reconnaissance phase, a deep learning model for generating stealthy attacks is initialized and trained in real-time by reconnoitering the compromised channels without launching any attacks. In the attacking phase, the malicious sensor measurements generated by the trained model are injected to replace the real measurements. Let the attacker conduct stealthy attacks by injecting a malicious program, then we outline two key parts of our framework for implementing stealthy attacks: a powerful model for generating stealthy attacks, and an effective method to train the model in real time. Therefore, in the remaining part of this section, we propose a GAN for generating stealthy attacks as well as a real-time learning method to train the stealthy attack GAN.

\subsection{Stealthy Attack GAN}
The stealthy attack GAN is composed by two deep learning models: a generator model for creating malicious sensor measurements and a discriminator model as a substitute anomaly detector to provide information for training the generator model. 

\subsubsection{Malicious Sensor Measurement Generator}
The objective of the generator model is to generate malicious sensor measurements which can achieve the predefined attack goals whilst bypassing the black box anomaly detector. Since the attacker does not have the full information of the physical process, the best strategy for the attacker is then to maximize the information he/she can utilize, which is the time series signals in the compromised channels, $\mathbf{X}_c$, to generate malicious sensor measurements.

Concretely, to utilize the information from the compromised channels, we define a sliding window:
\begin{equation*}
\mathbf{S}_c^{t}=\{ \mathbf{x}_c^{(t-l)}, \mathbf{x}_c^{(t-l+1)}, \ldots \mathbf{x}_c^{(t-1)} \}
\end{equation*}
which contains all the time series signals obtained from the compromised channels from time $t-l$ to $t-1$, where $l > 0$ is the length of the sliding window. Moreover, we also maintain another sliding windows $\tilde{\mathbf{S}}_c^{t}$ which only differs from $\mathbf{S}_c^{t}$ such that the previously generated malicious sensor measurements are injected, and their real values are replaced. Our generator will generate the next malicious sensor measurements $\tilde{\mathbf{y}}_c^{(t)}$ at time $t$ based on both $\mathbf{S}_c^{t}$ and $\tilde{\mathbf{S}}_c^{t}$. Intuitively, $\mathbf{S}_c^{t}$ captures the real system dynamics as well as providing the information relevant for achieving attack goals, $\tilde{\mathbf{S}}_c^{t}$ provides the related context for bypassing the black box anomaly detector with the consideration of the previously generated malicious measurements.

\begin{figure}
 \centering
 \includegraphics[width=0.99\columnwidth]{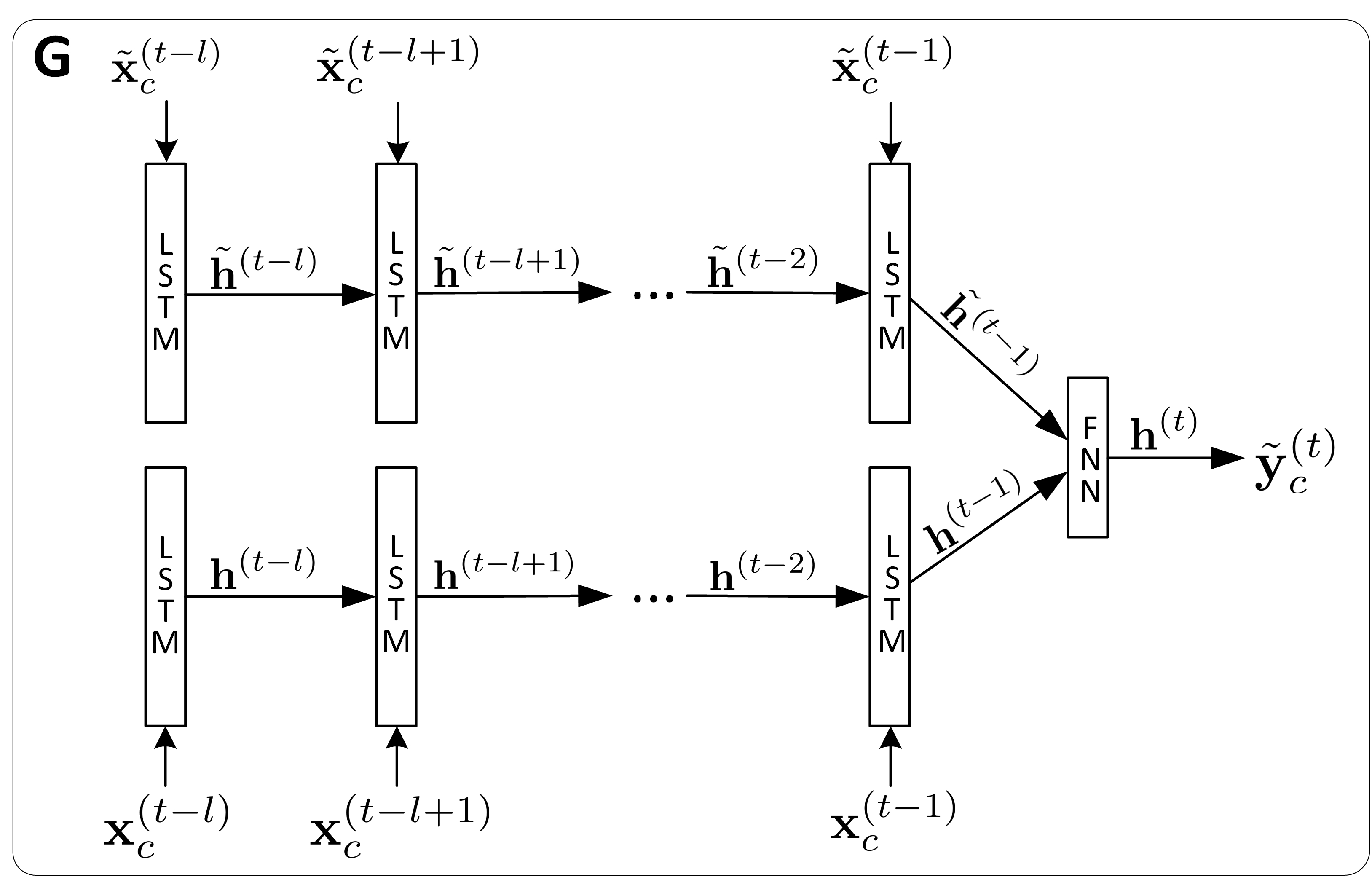}
 \caption{The generator model}
 \label{fig:generator}
\end{figure}

We approach malicious sensor measurement generation as a sequence learning problem, hence, we propose an LSTM-FNN as illustrated in Figure~\ref{fig:generator} to model our generator. Specifically, the two LSTMs read in the signals in the sliding window $\mathbf{S}_c^{t}$ and $\tilde{\mathbf{S}}_c^{t}$ separately, learn their temporal features, and then respectively encode them to hidden vectors $\mathbf{h}^{(t-1)}$ and $\tilde{\mathbf{h}}^{(t-1)}$, which provides the context for the generation of malicious sensor measurements. Then, the FNN will be used to learn their high dimensional features $\mathbf{h}^{(t)}$, and then output the malicious sensor measurements $\tilde{\mathbf{y}}_c^{(t)}$. Concretely, the model can be represented by the following equations:
\begin{eqnarray*}
\mathbf{h}^{(t-l)}&=&f(\mathbf{x}_c^{(t-l)}) \\
\mathbf{h}^{(t-i)}&=&f(\mathbf{x}_c^{(t-i)}, \mathbf{h}^{(t-i-1)}) \quad 1 \leq i \leq l-1 \\
\tilde{\mathbf{h}}^{(t-l)}&=&f(\tilde{\mathbf{x}}_c^{(t-l)}) \\
\tilde{\mathbf{h}}^{(t-i)}&=&f(\tilde{\mathbf{x}}_c^{(t-i)}, \tilde{\mathbf{h}}^{(t-i-1)}) \quad 1 \leq i \leq l-1 \\
\mathbf{h}^{(t)} &=& W_{h_1} \mathbf{h}^{(t-1)} +W_{h_2} \tilde{\mathbf{h}}^{(t-1)}  + \mathbf{b}_h \\
\tilde{\mathbf{y}}_c^{(t)} &=& W \mathbf{h}^{(t)} + \mathbf{b}
\end{eqnarray*} where the first four equations are used to iteratively encode $\mathbf{S}_c^{t}$ and $\tilde{\mathbf{S}}_c^{t}$ to hidden vectors $\mathbf{h}^{(t-1)}$ and $\tilde{\mathbf{h}}^{(t-1)}$ in which $f$ is a complex function as a short form of Equations~\ref{eq:dl:fg} to \ref{eq:dl:lstmh}; $W_{h_1}$, $W_{h_2}$ and $\mathbf{b}_h$ are the weight matrices and the bias vector to further encode the hidden vectors $\mathbf{h}^{(t-1)}$ and $\tilde{\mathbf{h}}^{(t-1)}$ for higher dimensional feature representation; $W$ and $b$ are the weight matrix and the bias vector respectively to decode $\mathbf{h}^{(t)}$ to the output malicious sensor measurements $\tilde{\mathbf{y}}_c^{(t)}$. For convenience, we represent the generator model as an overall function $G$:
\begin{equation} \label{eq:gen0}
\tilde{\mathbf{y}}_c^{(t)} = G(\mathbf{S}_c^{t}, \tilde{\mathbf{S}}_c^{t};\bth_G)
\end{equation}where $\bth_G$ denote the parameters of the model, $\mathbf{S}_c^{t}$, $\tilde{\mathbf{S}}_c^{t}$ and $\tilde{\mathbf{y}}_c^{(t)}$ are the inputs and output of the model, respectively.

Then, let $T$ be all the moments for generating malicious sensor measurements, $|T|$ be the size of $T$, to make the generated malicious measurements bypass the black box anomaly detector as well as achieving the attacker's goals is equivalent to optimize the generator model as follows:
\begin{align}\label{eq:gen1}
&\arg \min_{\bth_G} \: \frac{1}{|T|} \sum_{t \in T} \mathcal{F}(  G(\mathbf{S}_c^{t}, \tilde{\mathbf{S}}_c^{t};\bth_G) \mid \mathbf{X}^{t-1})\\
&\text{subject to }  \quad \tilde{y}_g^{(t)} \: \circled{$g$} \:  v_g  \quad \forall g\in \mathcal{G}, t \in T
\end{align}
From above, we can clearly see that in order to optimize the generator model, firstly, we have to use a substitute anomaly detection model to approximate the black box anomaly detection function $\mathcal{F}$ to provide information for the training of the generator.

\subsubsection{Substitute Anomaly Detector}
In order to provide information to optimize the generator of malicious sensor measurements, here we propose another neural network model to approximate the black box anomaly detector. Again, without the ability to access the entire time series $\mathbf{X}^{t-1}$, our strategy for defining the substitute anomaly detector is to utilize a sliding window $\hat{\mathbf{S}}_c^{t}$ ($\hat{\mathbf{S}}_c^{t}$ can either be $\mathbf{S}_c^{t}$ or $\tilde{\mathbf{S}}_c^{t}$ depending on whether malicious sensor measurements are injected in the previous time steps within the sliding window) to classify whether the sensor measurements at time $t$ are malicious or not.

 \begin{figure}
 \centering
 \includegraphics[width=0.99\columnwidth]{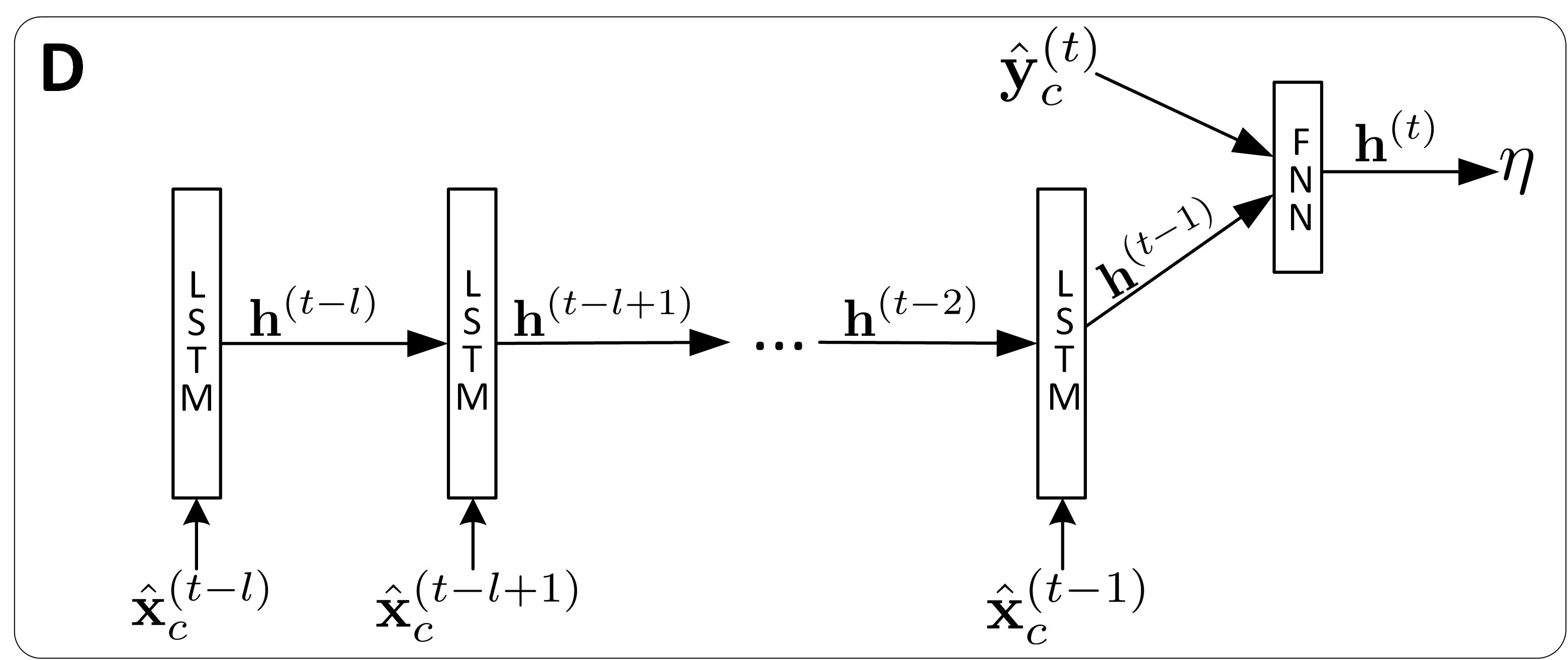}
 \caption{The substitute anomaly detector model}
 \label{fig:discriminator}
\end{figure}

Concretely, we employ an LSTM-FNN discriminator model whose architecture is illustrated in Figure~\ref{fig:discriminator} to model the substitute anomaly detector. The model consists of two parts: an LSTM which takes the sliding window $\hat{\mathbf{S}}_c^{t}$ as input, learns its temporal features, and then encodes them to a hidden vector $\mathbf{h}^{(t-1)}$; a FNN which takes $\mathbf{h}^{(t-1)}$ and the sensor measurements $\hat{\mathbf{y}}_c^{(t)}$ as input ($\hat{\mathbf{y}}_c^{(t)}$ can either be malicious measurements $\tilde{\mathbf{y}}_c^{(t)}$ or real measurements $\mathbf{y}_c^{(t)}$), encodes them to a hidden vector $\mathbf{h}^{(t)}$ for capturing nonlinear features, and then outputs a singular $\eta \in \mathbb{R}$ for classification. With a larger value of $\eta$, $\hat{\mathbf{y}}_c^{(t)}$ is more likely to be malicious. Note that in general, for a binary classification problem the sigmoid activation function will be used to transform the output $\eta$ to a probability $p \in (0,1)$ such that $p ={\frac {1}{1+e^{-\eta}}}$. However, since here we will use the training method inspired by WGAN, the sigmoid activation function is not used in our substitute anomaly detector.

With the discriminator model, the whole procedure for computing $\eta$ can be outlined by the following equations:
\begin{eqnarray*}
\mathbf{h}^{(t-l)}&=&f(\hat{\mathbf{x}}_c^{(t-l)}) \\
\mathbf{h}^{(t-i)}&=&f(\hat{\mathbf{x}}_c^{(t-i)}, \mathbf{h}^{(t-i-1)}) \quad 1 \leq i \leq l-1 \\
\mathbf{h}^{(t)} &=& W_h \mathbf{h}^{(t-1)} + W_y \hat{\mathbf{y}}_c^{(t)}  + \mathbf{b} \\
 \eta &=& \bm{\alpha} \: \mathbf{h}^{(t)} + \beta
\end{eqnarray*}
where the first two equations iteratively encode $\hat{\mathbf{S}}_c^{t}$ to a hidden vector $\mathbf{h}^{(t-1)}$ as similar with the generator model; the third equation encodes $\mathbf{h}^{(t-1)}$ and $\hat{\mathbf{y}}_c^{(t)}$ to a hidden vector $\mathbf{h}^{(t)}$ in which $W_h$ and $W_y$ are their weight matrices, $\mathbf{b}$ is a bias vector; the last equation decode $\mathbf{h}^{(t)}$ to the output singular $\eta$, in which $\bm{\alpha}$ is the weight vector, $\beta$ is a bias singular.

Again, for convenience, we represent the substitute anomaly detector as a function $D$:
\begin{equation*}
\eta = D( \hat{\mathbf{y}}_c^{(t)}, \hat{\mathbf{S}}_c^{t} ;\bth_D  )
\end{equation*}where $\bth_D$ is the parameters of the detector; $\hat{\mathbf{y}}_c^{(t)}$ and $\hat{\mathbf{S}}_c^{t}$ are the inputs; $\eta \in \mathbb{R}$ is the output. Then, the learning goal of $D$ is to output small values for real sensor measurements and large values for malicious sensor measurements in order to classify them. Hence, the optimization problem can be formulated as follows:
\begin{align}\label{eq:obd}
&\arg \min_{\bth_D} \:  \frac{1}{|T_1|} \sum_{t \in T_1} D\big( \mathbf{y}_c^{(t)} , \mathbf{S}_c^{t} ;\bth_D  \big) - \frac{1}{|T_2|} \sum_{t \in T_2} D\big( \tilde{\mathbf{y}}_c^{(t)} , \tilde{\mathbf{S}}_c^{t} ;\bth_D  \big)
\end{align}where $T_1$ and $T_2$ denote the moments for sampling real measurements and generated malicious measurements for training $D$, respectively; $\tilde{\mathbf{y}}_c^{(t)}$ and $\mathbf{y}_c^{(t)}$ represent a generated malicious measurement sample and a real measurement sample, respectively.

\subsubsection{The GAN}
With the substitute anomaly detector, then the optimization problem of the generator is equivalent to generating malicious sensor measurements which let the substitute anomaly detector output as smaller value as possible whilst achieving the attack goals. Specifically, we can replace the black box function $\mathcal{F}$ in Equation~\ref{eq:gen1} by the function $D$, then the optimization problem can be reformulated as follows:
\begin{align*}
&\arg \min_{\bth_G} \:  \frac{1}{|T|} \sum_{t \in T} D\big( G(\mathbf{S}_c^{t}, \tilde{\mathbf{S}}_c^{t};\bth_G) , \tilde{\mathbf{S}}_c^{t} ;\bth_D  \big) \\
&\text{subject to } \quad \tilde{y}_g^{(t)} \: \circled{$g$} \: v_g  \quad \forall g\in \mathcal{G}, t \in T
\end{align*}where we assume $\bth_D$ is fixed for the time being. Note that the above formulation requires that all the generated malicious sensor measurements can achieve the attack goals, which is, however, generally not feasible in practice (at some time points, if the attack goals are too ambitious, it is impossible to generate such measurements both bypassing the anomaly detector and achieving attack goals). As a result, we relax the optimization problem to allow the attack goals to fail in some time points, but we pay a cost from each failed cases. To implement this, we introduce slack variables $\xi_g^{(t)} \in \mathbb{R}^{\geq 0}$. Specifically, a non-zero value for $\xi_g^{(t)}$ allows a generated sensor measurement $\tilde{y}_g^{(t)}$ to not satisfy an attack goal $g$ at a cost proportional to the value of $\xi_g^{(t)}$.

With slack variables, the formulation of the optimization problem becomes:
\begin{align} \label{eq:op1}
&\arg \min_{\bth_G} \: \frac{1}{|T|} \sum_{t \in T} [ D\big( G(\mathbf{S}_c^{t}, \tilde{\mathbf{S}}_c^{t} ;\bth_G) , \tilde{\mathbf{S}}_c^{t} ;\bth_D  \big) +   \sum_{g \in \mathcal{G}} \lambda_g \xi_g^{(t)} ] \\\label{eq:op2}
&\text{subject to } \quad  \tilde{y}_g^{(t)} \: \circled{$g$} \:  v_g \pm \xi_g^{(t)} \quad \forall g\in \mathcal{G}, t \in T
\end{align}where the second equation means we achieve attacks goals by either adding or subtracting slack variables $\xi_g^{(t)}$; $\lambda_g \in \mathbb{R}^{>0}$ in the first equation is a hyperparameter which controls the trade-off between the probability of bypassing the substitute anomaly detector and the distance to achieve the attack goal: as $\lambda_g$ becomes larger, the generator model is more willing to generate sensor measurements to achieve attack goals; when $\lambda_g$ is small, the model is more likely to generate sensor measurements which can bypass the anomaly detector, but the attack goals are not strictly satisfied. In our case, $\lambda_g$ is always set to a small value as we should always prioritize the generator's ability to bypass the anomaly detector.

More importantly, we can always find such slack variables to satisfy the constraints in Equation~\ref{eq:op2}, and the value of slack variables can be obtained by the following equation:
\begin{align*}
\xi_g^{(t)} = \begin{cases}
    \max (0, v_g-\tilde{y}_g^{(t)})      & \quad \text{if } \circled{$g$} \in \{ >,\geq \}\\
    |v_g-\tilde{y}_g^{(t)}| &  \quad \text{if } \circled{$g$} \in \{ = \}\\
    \max (0,\tilde{y}_g^{(t)}-v_g) &  \quad \text{if } \circled{$g$} \in \{ <,\leq \}\\
  \end{cases}
\quad  \forall g\in \mathcal{G},  t \in T
\end{align*} Replacing the above equation into Equation~\ref{eq:op1}, we can finally remove the constraints in Equation~\ref{eq:op2}, then the constrained optimization problem for the generator is converted to an unconstrained optimization problem over $\bth_G$.

\begin{figure}
 \centering
 \includegraphics[width=0.99\columnwidth]{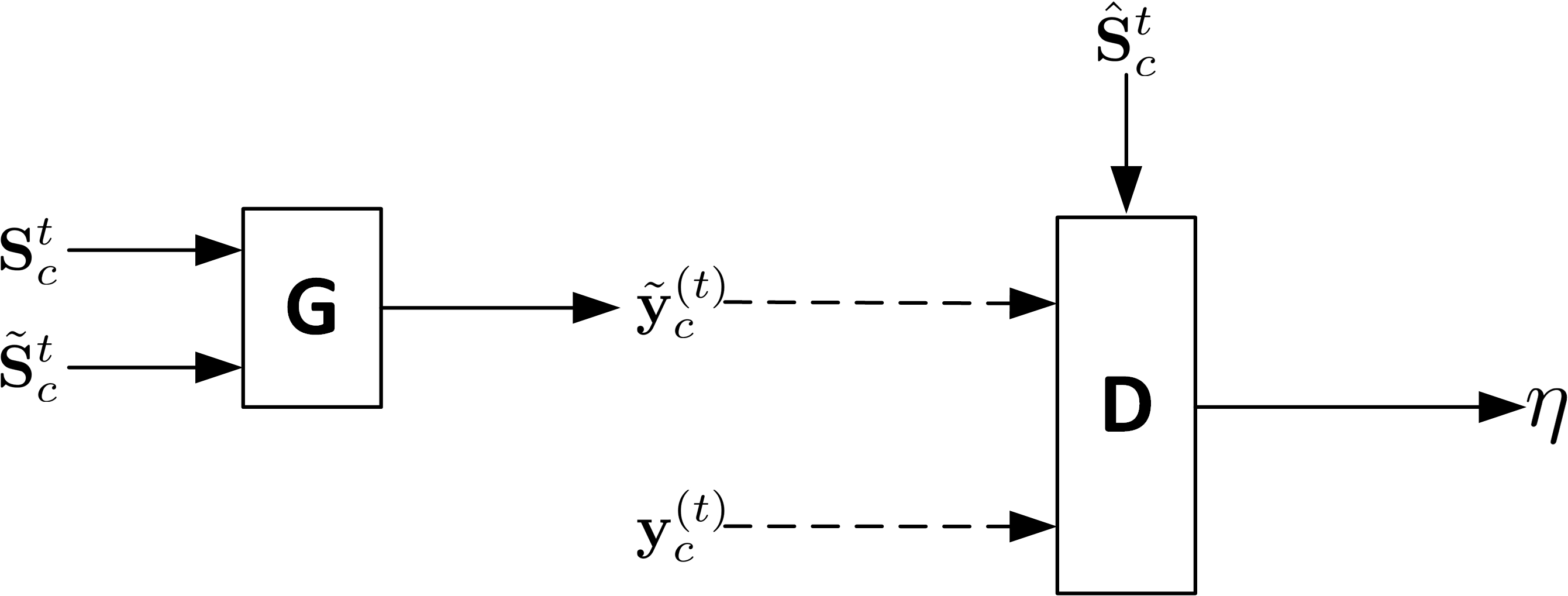}
 \caption{The architecture of the stealthy attack GAN}
 \label{fig:AttackGAN}
\end{figure}

With the completion of the definition of the generator and the substitute anomaly detector, we illustrate the architecture of the stealthy attack GAN in Figure~\ref{fig:AttackGAN}, in which dashed arrows indicate optional data flow.

\subsection{Real-time Learning Method}
Here we present the learning method to train the stealthy attack GAN. Specifically, the basic principle of our learning method follows the training principle of WGAN, which is to train the generator and the substitute anomaly detector iteratively to play an adversarial game until the generated malicious sensor measurements cannot be distinguished from real measurements by the substitute anomaly detector. However, since it is generally not feasible for the attacker to collect the intercepted time series signals in a repository and then train the stealthy attack GAN in an offline mode, we propose a real-time learning method to train the stealthy attack GAN. The whole procedure of the learning method is illustrated in Algorithm~\ref{alg:training}.

\begin{algorithm}[H]
  \caption{The real-time learning method of the stealthy attack GAN}
  \begin{algorithmic}[1]
  	\REQUIRE the learning rate $\alpha=0.00005$, the clipping parameter $c=0.01$, the number of time steps for training the substitute anomaly detector per time step for training the generator $n=5$.
  	\REQUIRE a sliding window $\mathbf{S}^t_c$ with real sensor measurements, and another sliding window $\tilde{\mathbf{S}}^t_c$ with generated malicious sensor measurements.
  	\REQUIRE the length of sliding window $l$, the total time steps for learning $N$, the trade-off hyperparameter $\lambda_g, \forall g \in \mathcal{G}$, the probability $\upsilon$ for reseting $\tilde{\mathbf{S}}^t_c$ with $\mathbf{S}^t_c$
  	\REQUIRE the initial parameters of the substitute detector $\bth_D^0$, the initial parameters of the generator $\bth_G^0$, the initial sliding windows $\tilde{\mathbf{S}}^l_c=\mathbf{S}^l_c$
     \FOR{$t=l+1,l+2,\ldots, l+N$}
     \STATE{Generate a random number $\gamma$ uniformly distributed in $(0,1)$}
     \IF{$\gamma < \upsilon$}
     \STATE{Set $\tilde{\mathbf{S}}^t_c = \mathbf{S}^t_c$  }
     \ENDIF
      \IF{$t \mod (n+1) \neq 0$}
     	\STATE{Generate malicious measurements: \begin{equation*}
     	\tilde{\mathbf{y}}_c^{(t)} = G( \mathbf{S}^t_c, \tilde{\mathbf{S}}_c^{t};\bth_G) 
     	\end{equation*} }
     	\STATE{$\delta_{\bth_D} \leftarrow \nabla_{\bth_D} [ D(  \mathbf{y}_c^{(t)}, \mathbf{S}^t_c ;\bth_D  ) -  D( \tilde{\mathbf{y}}_c^{(t)},\tilde{\mathbf{S}}^t_c ;\bth_D) ] $}
     	\STATE{$\bth_D \leftarrow \bth_D + \alpha \cdot RMSProp(\bth_D, \delta_{\bth_D})$}
     	\STATE{$\bth_D \leftarrow clip(\bth_D , -c, c)$}
     	\STATE{Reset $\mathbf{S}^t_c$ and $\tilde{\mathbf{S}}^t_c$}
     \ENDIF
     \IF{$t \mod (n+1) = 0$}
     \STATE{fix $\bth_D$}
     \STATE{\begin{small} $\delta_{\bth_G} \leftarrow -\nabla_{\bth_G}  [ D\big( G(\mathbf{S}^t_c, \tilde{\mathbf{S}}^t_c;\bth_G) , \tilde{\mathbf{S}}^t_c ;\bth_D  \big) +   \sum_{g \in \mathcal{G}} \lambda_g \xi_g^{(t)} ]$ \end{small} }
	\STATE{$\bth_G \leftarrow \bth_G + \alpha \cdot RMSProp(\bth_G, \delta_{\bth_G})$}
	\STATE{Reset $\mathbf{S}^t_c$ and $\tilde{\mathbf{S}}^t_c$}
	\ENDIF
      \ENDFOR
  \end{algorithmic}
   \label{alg:training}
\end{algorithm}

Specifically, our learning method only requires to maintain two sliding windows $\mathbf{S}_c^{t}$ and $\tilde{\mathbf{S}}_c^{t}$ for the time series signals in the compromised channels with the real sensor measurements and the corresponding generated malicious sensor measurements. At each time step, either the substitute anomaly detector is trained to minimize the objective function in Equation~\ref{eq:obd} by using the gradient at the current sample which contains the real as well as the generated malicious measurements for the time step (Steps 7 to 9), or the generator is trained to minimize the objective function in Equation~\ref{eq:op1} by the gradient at the current sample in which the two sliding windows are used to generate malicious measurements at the time step to fool the substitute anomaly detector as well as achieving the attack goals (Steps 14 to 16). At each time step, the two sliding windows are reseted such that new signals at time $t$ will be appended, and the expired signals at time $t-l$ will be removed (Steps 11 and 17). Moreover, we also periodically reset $\tilde{\mathbf{S}}_c^{t}=\mathbf{S}_c^{t}$ at a rate specified by $\upsilon$ to start a new cycle for continuously injecting malicious measurements (Steps 2 to 5). Note that reseting $\tilde{\mathbf{S}}_c^{t}=\mathbf{S}_c^{t}$ periodically is rather important since this gives the chance for the stealthy attack GAN to properly learn how to initiate the attacks.

After the training of the stealthy attack GAN has been finished, the attacker can generate malicious sensor measurements for each time step at the attacking phase using Equation~\ref{eq:gen0}, and then inject them to the corresponding compromised channels to replace the real measurements for achieving their attack goals.

\section{Gas Pipeline Case Study}
In this section, we show a case study in which we conduct stealthy attacks in a laboratory-scale gas pipeline system using our framework. Specifically, the system in our case study consists of a small airtight pipeline connected to a pressure meter, a pump, and a solenoid-controlled relief valve. A SCADA system is deployed to control the air pressure in the pipeline, which contains a PLC, a sensor for pressure measurement, and several actuators. Our attack goal is to constantly fool the PLC with a malicious pressure measurement which is smaller than its real value with different scales, which can potentially lead to the explosion of the gas pipeline. All the experiments in the section are done by exploiting a public dataset \cite{morrisindustrial} which records the network traffic data log captured from the gas pipeline SCADA system.

\subsection{Gas Pipeline Dataset}
The gas pipeline dataset for training and testing our stealthy attack GAN consists of the sensor measurements and control commands extracted every two seconds from normal network packages in a gas pipeline network traffic data log originally described in \cite{morrisindustrial}. In total, 68,803 time series signals are collected. The detailed description of the extracted features for each signal is listed in Table~\ref{tab:gasdata}. Specifically, the gas pipeline includes a system control mode with 3 states; off, manual control, or automatic control. In off mode, the pump is forced to the off state to allow it to cool down. In manual mode, a HMI can be used to manually change the pump state and control the relief valve state by opening or closing the solenoid. In automatic mode, the control scheme determines which mechanism is used to regulate the pressure set point: either by turning a pump on or off or by opening and closing a relief valve using a solenoid. Moreover, a proportional integral derivative (PID) controller is used to control the pump or solenoid depending upon the control scheme chosen. Six PID control parameters can be set, which are pressure set point, gain, reset rate, rate, dead band, and cycle time.
\begin{table}
\centering
\begin{small}

\begin{tabular}{|l|l|}
\hline
\textbf{Feature}            & \textbf{Description}                                                   \\ \hline
\textit{setpoint}             & The pressure set point  \\ \hline
\textit{gain}                 & PID gain                                                               \\ \hline
\textit{reset rate}           & PID reset rate                                                         \\ \hline
\textit{deadband}             & PID dead band                                                          \\ \hline
\textit{cycle time}           & PID cycle time                                                         \\ \hline
\textit{rate}                 & PID rate                                                               \\ \hline
\textit{system mode}          & automatic (2), manual (1) or off (0)             \\ \hline
\textit{control scheme}       & Either pump (0) or solenoid (1)                                        \\ \hline
\textit{pump}                 &\pbox{20cm}{\vspace{1mm} Pump control -- open (1) or off (0)  \vspace{-0.5mm} \\  only for manual mode \vspace{1mm}}  \\ \hline
\textit{solenoid}                 &\pbox{20cm}{\vspace{1mm} Valve control -- open (1) or closed (0)  \vspace{-0.5mm} \\only for manual mode.  \vspace{1mm}}  \\ \hline
\pbox{20cm}{\vspace{1mm} \emph{      \textbf{pressure}} \vspace{-0.5mm} \\  \emph{\textbf{measurement}} \vspace{1mm}}  & Pressure measurement   \\ \hline
\end{tabular}

\caption{Extracted sensor measurements (in bold text) and control commands from the gas pipeline dataset \cite{morrisindustrial}}
\label{tab:gasdata}
\end{small}
\end{table}

\subsection{Experiment Setup}
In our experiments, we split the time-series dataset into two slices. The first slice which contains $4/5$ of the data is used for the reconnaissance phase to train our stealthy attack GAN. The other slice is used for the attacking phase. Since the gas pipeline dataset is relatively small, we go through the first time-series slice for 50 passes to properly train the stealthy attack GAN using our real-time learning method.

\subsubsection{Baseline Anomaly Detector}
For the baseline anomaly detector, three predictive models, which are the AR model, the LDS model, and the LSTM model as described in Section~\ref{sec:predictive} are fitted to minimize the mean square error between the predicted pressure measurements and their real values using cross validation on the first time-series slice. We pick the model with the best prediction accuracy, which is the LSTM model (with $10$ previous signals are used for the model inputs) as our baseline predictive model. Both the residual errors and the CUSUM statistics as described in Section~\ref{sec:detect} are used to detect anomalies.


\subsubsection{Attack Scenarios}
The normal air pressure range for the gas pipeline system is about [0,40]. Let $y_g^{(t)}$ be the real pressure measurement at time $t$, we investigate two attack goals, which are to inject malicious measurement $\tilde{y}_g^{(t)}$ such that $\tilde{y}_g^{(t)}=\max(y_g^{(t)}-4,0)$ and $\tilde{y}_g^{(t)}=\max(y_g^{(t)}-8,0)$. Specifically, the above attack goals require the malicious measurements being 4 or 8 units smaller than their real values (but not beyond the normal pressure range). Furthermore, we also consider two cases, in which the attacker can only compromise the PLC-sensor channel, and the attacker can compromise all the PLC-sensor, PLC-actuator channels in the gas pipeline system. To summarize, we consider four attack scenarios with different attacker's goal and ability assumptions as illustrated in Table~\ref{tab:attack-scenario}.

\begin{table}
\centering
\begin{small}
\scalebox{0.8}{
\begin{tabular}{|c|c|c|c|}
\hline
\multicolumn{2}{|c|}{\multirow{2}{*}{}}                                                                                                                         & \multicolumn{2}{c|}{\textbf{Attack Goal}}                                                  \\ \cline{3-4}
\multicolumn{2}{|c|}{}                                                                                                                                          & $\tilde{y}_g^{(t)} = max(y_g^{(t)}-4, 0)$ & $\tilde{y}_g^{(t)} = max(y_g^{(t)}-8, 0)$ \\ \hline
\multirow{2}{*}{\begin{tabular}[c]{@{}c@{}}\textbf{Attacker's} \\ \textbf{Abilities}\end{tabular}} & \begin{tabular}[c]{@{}c@{}}PLC-Sensor \\ channel\\ Compromised\end{tabular} & Attack Scenario 1                         & Attack Scenario 2                         \\ \cline{2-4}
                                                                                 & \begin{tabular}[c]{@{}c@{}}All channels\\ Compromised\end{tabular}           & Attack Scenario 3                         & Attack Scenario 4                         \\ \hline
\end{tabular}
}\caption{Attack Scenarios of the Gas Pipeline Case Study}

\label{tab:attack-scenario}
\end{small}
\end{table}

\subsubsection{Feature Processing}
We normalize all continuous features into range $[0,1]$ by min-max scaling. Specifically, let $x_i^{(t)}$ be a continuous feature, e.g., the setpoint, at time $t$, $\min(x_i)$ and $\max (x_i)$ respectively be the minimal and maximal value of the feature, we covert $x_i^{(t)} = \frac{x_i^{(t)}-min(x_i)}{max(x_i)-min(x_i)}$. All the categorical features (system mode, control scheme, pump, solenoid), are one-hot encoded, e.g., automatic, manual, off system modes are encoded to [1,0,0], [0,1,0], [0,0,1], respectively.

\subsubsection{Model Parameters and Memory Cost}
We explicitly set the number of units in the LSTM layers and the FNN layers of the stealthy attack GAN to four and two times of the dimension of the processed signals, respectively. The length of sliding window $l$ is set to 10, which is equal to the number of previous signals used in the baseline LSTM predictive model. The trade-off hyperparameter $\lambda_g$ is set to $0.01$ as we find it achieves the best balance between bypassing anomaly detectors and achieving attack goals. The probability $\upsilon$ for reseting $\tilde{\mathbf{S}}^t_c$ with $\mathbf{S}^t_c$ is also set to $0.01$. The memory costs for the stealthy attack GANs are about 40 kB for the first two attack scenarios, 160 kB for the other two attack scenarios implemented using the Keras library \cite{chollet2015keras}.

\begin{figure*}
  \begin{subfigure}
  \centering
  \includegraphics[width=0.24\linewidth]{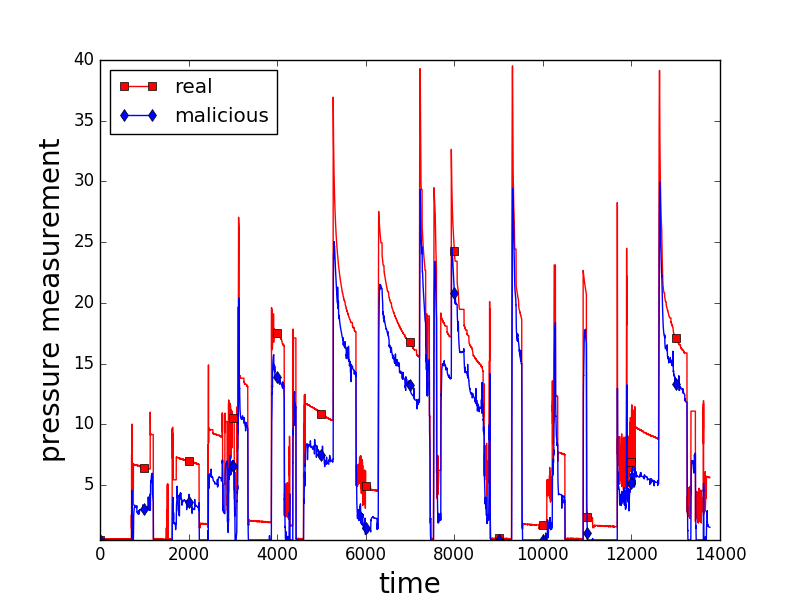}
\end{subfigure}%
\begin{subfigure}
  \centering
  \includegraphics[width=0.24\linewidth]{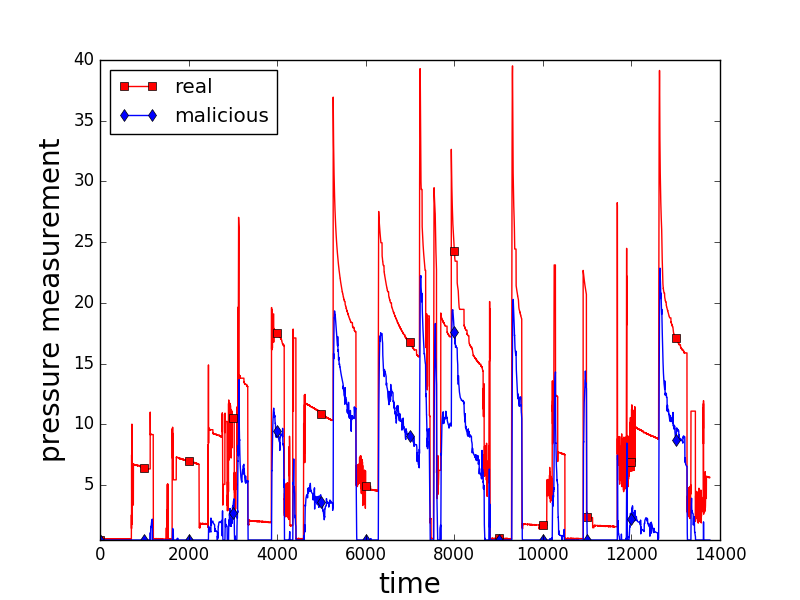}
\end{subfigure}
\begin{subfigure}
  \centering
  \includegraphics[width=0.24\linewidth]{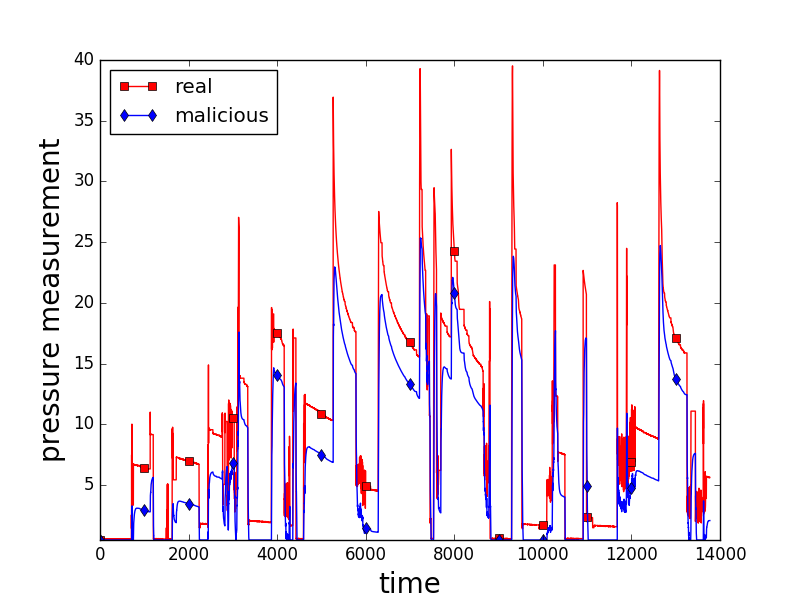}
\end{subfigure}
\begin{subfigure}
  \centering
  \includegraphics[width=0.24\linewidth]{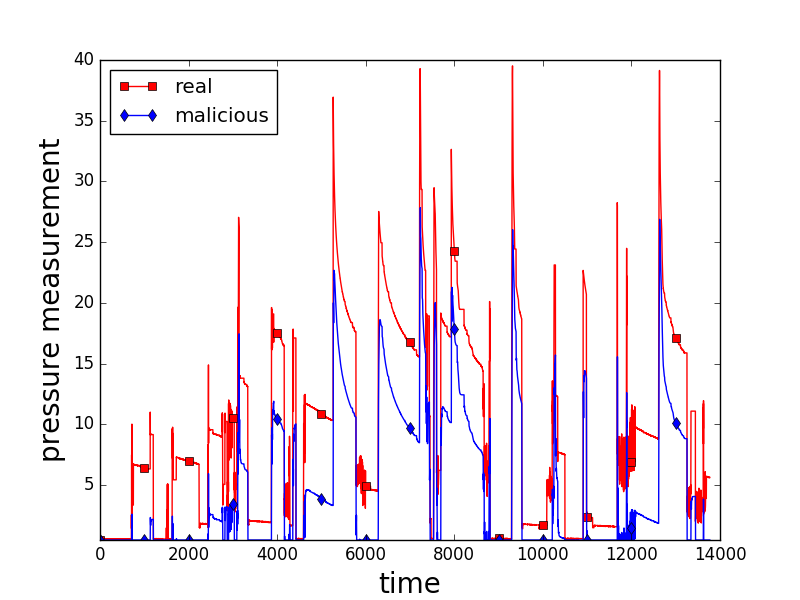}
\end{subfigure}
\caption{The generated malicious pressure measurement trace in the attacking phase compared with their real values. (Scenario 1 to 4 from left to right)}
  \label{fig:pressure}
\end{figure*}

\begin{figure*}
  \begin{subfigure}
  \centering
  \includegraphics[width=0.24\linewidth]{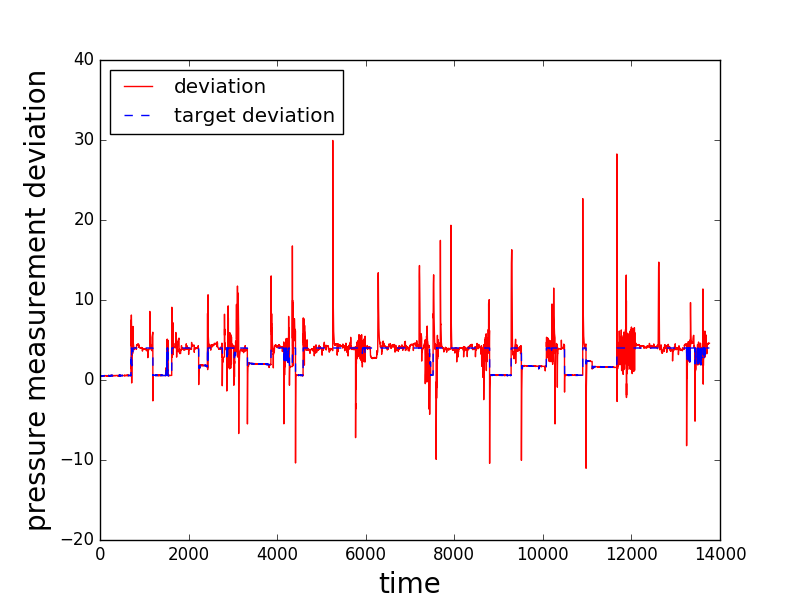}
\end{subfigure}%
\begin{subfigure}
  \centering
  \includegraphics[width=0.24\linewidth]{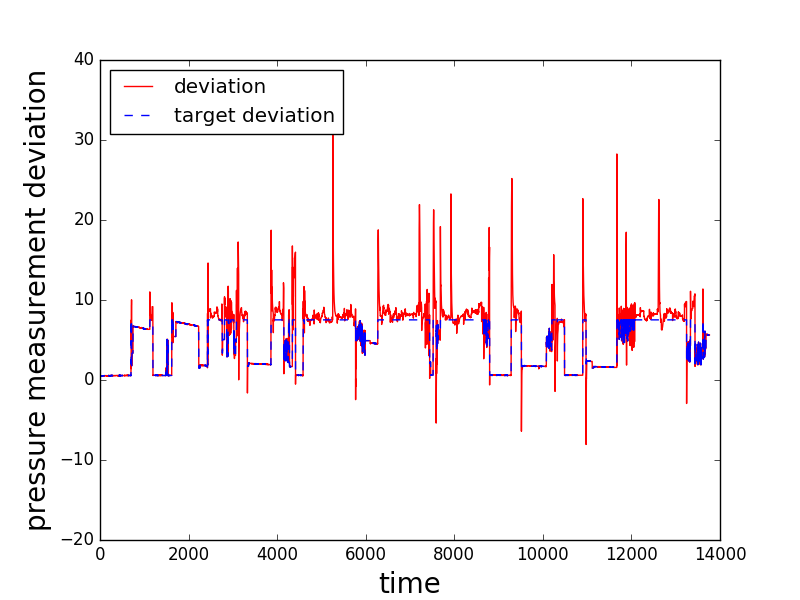}
\end{subfigure}
\begin{subfigure}
  \centering
  \includegraphics[width=0.24\linewidth]{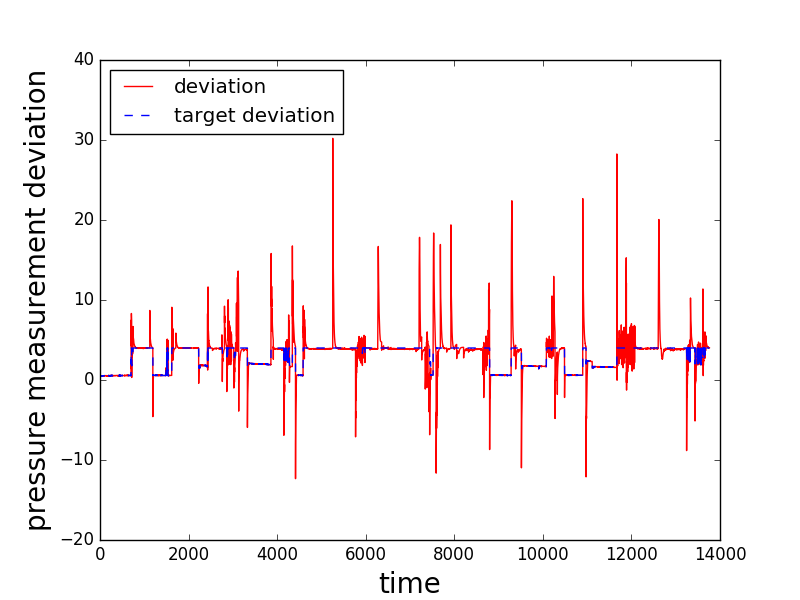}
\end{subfigure}
\begin{subfigure}
  \centering
  \includegraphics[width=0.24\linewidth]{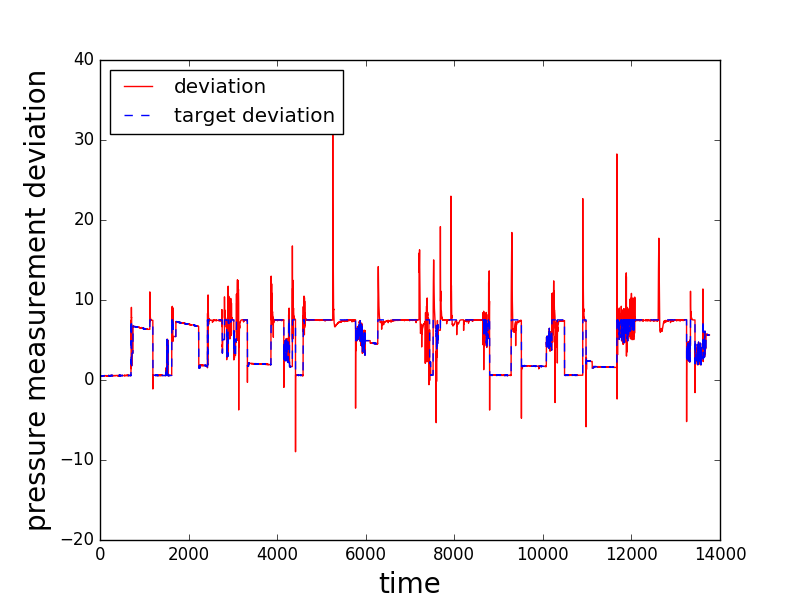}
\end{subfigure}
\caption{The deviation between the generated malicious pressure measurements and their real values at each time point in the attacking phase compared with the target deviation as specified by the attack goals (Scenario 1 to 4 from left to right)}
  \label{fig:deviation}
\end{figure*}

\begin{figure}
  \centering
  \includegraphics[width=0.99\linewidth]{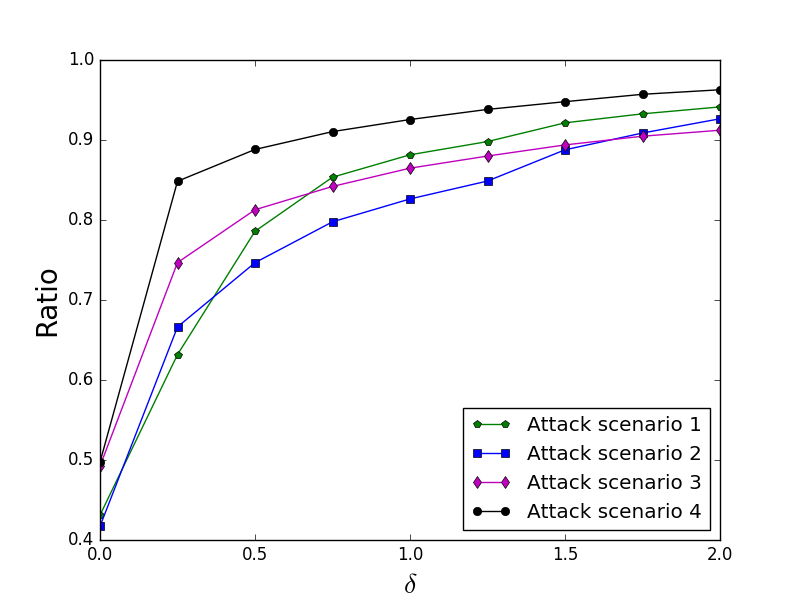}
\caption{The ratio of time points at which the attack goal is achieved in the attacking phase with different value of $\delta$}
  \label{fig:goal}
\end{figure}

\subsection{Results and Evaluation}
For each attack scenario, we train the corresponding stealthy attack GAN using our real-time learning method during the reconnaissance phase, after which we start to generate malicious pressure measurement for each time point in the attacking phase. To evaluate the quality of the malicious pressure measurements generated by our stealthy attack GAN, we show the trace of malicious pressure measurements during the attacking phase in the four attack scenarios compared with the trace of the real values in Figure~\ref{fig:pressure}. As can be seen from the figure, the generated malicious pressure measurements successfully capture the trend of the real trace. The trace of malicious measurements in the first two scenarios exhibit a minor fluctuation behavior as only the PLC-sensor channel can be compromised, thus the generated malicious measurements are more noisy than the counterpart in the other two scenarios.

Furthermore, we also illustrate the deviation between the generated malicious pressure measurements and their real values at each time point in the attacking phase compared with the target deviation as specified by the attack goals in Figure~\ref{fig:deviation}. We can see that at most time points, the deviation is very close with the target deviation, which means that the attack goals are well achieved at these time points. We are aware that there are a few specific time points at which the deviations are rather far away with the target deviation. They are mostly caused by the noisy behaviour of air pressure in the gas pipeline system induced by human operations from HMI. The pressure measurements at these time points are hard to capture, and thus can be seen as outliers for the deployed anomaly detector (note that in reality, human operations are much less often than here in the gas pipeline dataset). As a result, our stealthy attack GAN has to sacrifice the attack goal in order to prioritize capturing the behaviour of the anomaly detector so as to bypass it at these time points. Nevertheless, we can also observe that the stealthy attack GAN can adjust the deviation to the targeted value quickly after these outliers. Furthermore, to quantitatively evaluate the quality of the generated malicious measurements, we consider the attack goal is achieved at a time point $t$ if $|d_{real}^t - d_{target}^t| \leq \delta$, where $d_{real}^t$ and $d_{target}^t$ respectively denote the real deviation and the target deviation between the malicious pressure measurement and its real value at time point $t$, $\delta$ is a threshold value. In Figure~\ref{fig:goal}, we plot the ratio of time points at which the attack goal is achieved with different values of $\delta$ in each attacking scenario. We can observe that at around $50\%$ of time points, the deviation caused by our attacks \emph{exactly} matches the target deviation ($\delta=0$) if  both the PLC-sensor and the PLC-actuator channels are compromised. The figure is slightly lower (about $40\%$) if only the PLC-sensor channel is compromised due to the fluctuation behavior as illustrated in Figure~\ref{fig:pressure}. And we can see that the ratios in all the attack scenarios become close when we increase $\delta$ by which the standard for achieving attack goals is relaxed. This means the generated malicious pressure measurements by our stealthy attack GAN are rather robust in achieving attack goals even with different scales of target deviation and different knowledge of the system dynamics. Moreover, we can observe that even with $\delta=1$, attack goals are achieved more than $80\%$ time points in all attacking scenarios, this figure increases to more than $90\%$ when $\delta=2$, which indicates the high quality of malicious pressure measurements generated by our stealthy attack GAN for achieving the targeted deviations.

\begin{figure}
  \centering
  \includegraphics[width=0.99\linewidth]{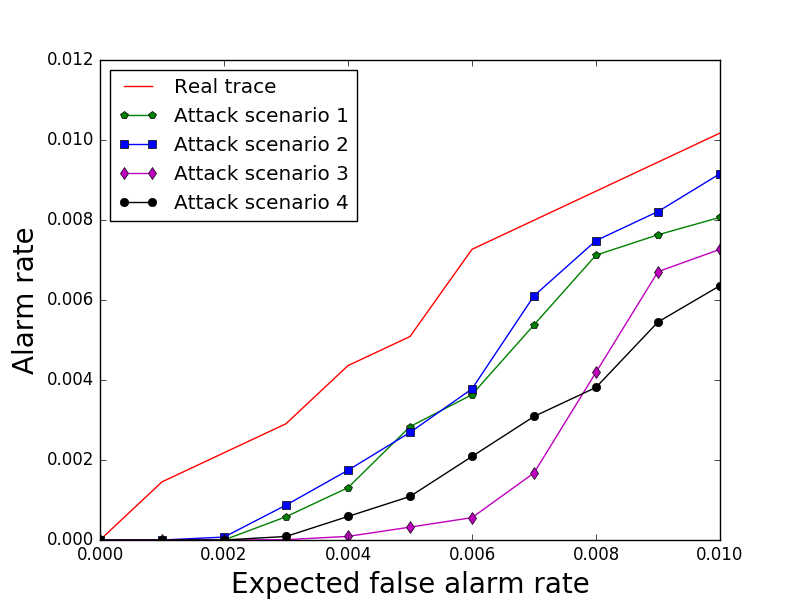}
\caption{The values of alarm rates in the four attack scenarios compared with the corresponding value for the real pressure measurement trace with different values of expected false alarm rate}
  \label{fig:residual}
\end{figure}

\begin{figure}
  \centering
  \includegraphics[width=0.9\linewidth]{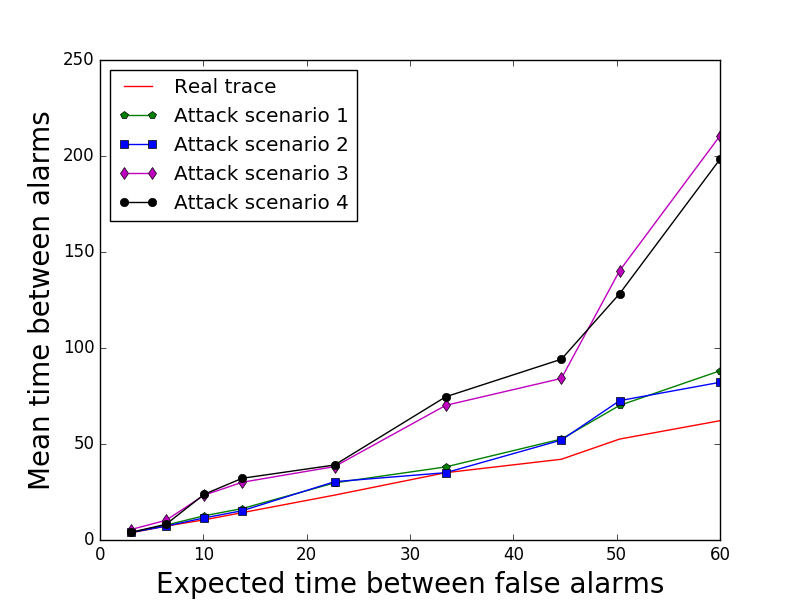}
\caption{The values of mean time between alarms (measured in minutes) in the four attack scenarios compared with the corresponding value for the real pressure measurement trace with different values of expected time between false alarms}
  \label{fig:cusum}
\end{figure}

To demonstrate the ability of the generated malicious pressure measurements to bypass the black box anomaly detector. We use both the residual error and the CUSUM statistic-based methods to detect the injected malicious pressure measurements in all the four attack scenarios. Specifically, we tune the value of $\tau$ for the residual error-based method to get different values of expected false alarm rate based on the training dataset. The value of $\tau$ for the CUSUM statistic-based method is also tunned to get different values of expected time between false alarms. In Figure~\ref{fig:residual} and~\ref{fig:cusum}, we plot the values of alarm rates with different values of expected false alarm rate and the value of mean time between alarms (measured in minutes) with different values of expected time between false alarms in the four attack scenarios, respectively, comparing with the corresponding values for the real pressure measurement trace. From Figure~\ref{fig:residual}, we can observe that in all cases, the alarm rates for our injected malicious pressure measurements are lower than the corresponding value for the measurements in the real trace. This indicates that it is extremely inefficient and costly (with respect to the expected false alarm rate) to detect our generated stealthy attacks using the residual error-based detection method. Additionally, we can also see that with the expected false alarm rate being less than $0.002$ ($0.002$ expected false alarm rate means that it is expected that 2 out of 1000 detections will generate a false alarm), it is almost impossible to detect our stealthy attacks in all scenarios. Furthermore, from Figure~\ref{fig:cusum}, we also observe that, in all cases, the mean time between alarms for our injected malicious pressure measurements are higher than the corresponding value for the measurements in the real trace. This reflects that it is also very inefficient to use the CUSUM statistic-based method to detect our stealthy attacks. The reason for all the above phenomena is that instead of minimizing the mean square error between the predicted pressure measurements and their real values like traditional predictive models, our stealthy attack GAN is configured to \emph{directly learn how to bypass the black box anomaly detector}, thus can continuously generate malicious sensor measurements with smaller residual errors even compared with the real trace, making them extremely difficult to reveal by both the residual error and the CUSUM statistic-based detection methods. Lastly, we also observe from the results for the first two attack scenarios that only compromising the PLC-sensor channel can still generate high-quality stealthy attacks. We believe this is because there exist unstable delays to reflect the control commands from sensor measurements, which allows the stealthy attack GAN to learn the control commands solely from the previous real sensor measurements in the sliding window $\mathbf{S}_c^{t}$, making the generated malicious sensor measurements more likely to bypass the black box anomaly detector.



\section{Water Treatment System Case Study}\label{sec:swat}
In this section, we present another case study in which we conduct stealthy attacks using our framework onto a Secure Water Treatment (SWaT) testbed using a simulation approach. Specifically, the SWaT testbed \cite{gohdataset} is a scaled down water treatment plant which has a six-stage filtration process to purify raw water. The general description of each stage is as follows: In the first stage, raw water is taken in and stored in a tank. It is then passed to the second stage for pretreatment process, where the conductivity, pH, and Oxidation Reduction Potential (ORP) are measured to determine whether chemical dosing is performed to maintain the water quality within acceptable limits. In the third stage, the Ultra Filtration (UF) system will remove undesirable materials by using fine filtration membranes. This is followed by the fourth stage, where the remaining chlorines are destroyed in the Dechlorinization process using Ultraviolet lamps. Subsequently, the water is pumped into the Reverse Osmosis (RO) system to reduce inorganic impurities in the fifth stage. In the last stage, the clean water from the RO system is stored and ready for distribution. A SCADA system with 6 PLCs, 24 sensors and 27 actuators is deployed to control the testbed. Our attacks focus on the second stage, where we inject malicious pH and ORP measurements which are larger and smaller than their real values simultaneously to change the quality of produced water.

\subsection{Dataset and Experiment Setup}
The water treatment dataset for training and testing our stealthy attack GAN consists of 51 sensor measurements and control commands extracted every second from the SWaT dataset \cite{goh2017anomaly}. In total, 496,800 signals for regular operation are collected. Since we focus on conducting stealthy attacks in the second stage of the water treatment process, we only illustrate the related sensor measurements and control commands for the second stage in Table~\ref{tab:waterdata}.

\begin{table}
\centering
\begin{small}
\begin{tabular}{|l|l|}
\hline
\textbf{Feature}            & \textbf{Description}                                                   \\ \hline
\textbf{\textit{AIT201}}                 &\pbox{20cm}{\vspace{1mm} Conductivity analyzer \vspace{-0.5mm} \\ measures NaCl level  \vspace{1mm}}  \\ \hline
\textbf{\textit{AIT202}}                &\pbox{20cm}{\vspace{1mm} pH analyzer \vspace{-0.5mm} \\ measures HCl level  \vspace{1mm}}  \\ \hline
\textbf{\textit{AIT203}}               &\pbox{20cm}{\vspace{1mm} ORP analyzer \vspace{-0.5mm} \\ measures NaOCl level  \vspace{1mm}}  \\ \hline
\textbf{\textit{FIT201}}             &  Flow transmitter for dosing pumps \\ \hline
\textit{P101}                 &\pbox{20cm}{\vspace{1mm} Raw water tank pump state \vspace{-0.5mm} \\ control water from tank to the second stage  \vspace{1mm}}  \\\hline
\textit{MV201}                 &\pbox{20cm}{\vspace{1mm} Motorized valve state \vspace{-0.5mm} \\ control water flow to the third stage  \vspace{1mm}}  \\ \hline

\textit{P201}           & NaCl dosing pump state \\ \hline
\textit{P203}             & HCl dosing pump state \\ \hline
\textit{P205}           & NaOCl dosing pump state \\ \hline
\end{tabular}
\caption{Extracted sensor measurements (in bold text) and control commands from the water treatment dataset \cite{goh2017anomaly}}
\label{tab:waterdata}
\end{small}
\end{table}

Again, we split the time-series dataset into two slices. The first slice for training the stealthy attack GAN contains $4/5$ of the data. The other slice is used for the attacking phase. We also use the LSTM predictive model (as it achieves the best prediction accuracy) together with the residual error as well as the CUSUM statistic-based method for our baseline anomaly detector. For feature processing, we also normalize all the continuous features in the dataset using min-max scaling. All categorical features are one-hot encoded. All the model parameters are also set with the same rule as in the previous case study.

Here we set the attack goal to simultaneously send malicious HCl and NaOCl measurements to their corresponding PLCs whose values are $0.1$ (after normalization) higher and lower than their real values, respectively, but not beyond the normal range. Specifically, let $y_{g_1}^{(t)}, y_{g_2}^{(t)}$ be the real HCl and NaOCl level measurements at time $t$, the attacker requires the malicious measurements sent to the PLCs as $\tilde{y}_{g_1}^{(t)}\geq \min( y_{g_1}^{(t)} + 0.1, 1 )$ and $\tilde{y}_{g_2}^{(t)}\leq \max(\tilde{y}_{g_2}^{(t)}-0.1, 0)$. Furthermore, we consider two attack scenarios. In the first scenario, the attacker can only compromise the PLC-AIT202, PLC-AIT203 channels. In the second scenario, the attacker is able to compromise all the PLC-sensor, PLC-actuator channels in the stage. The corresponding memory costs of the stealthy attack GANs in the two scenarios are about 44 kB and 154 kB, respectively.

\subsection{Simulation and Results}
We test our framework for conducting stealthy attacks in this case study using a simulation approach. Specifically, in the SWaT testbed, the HCl and NaOCl dosing pumps are always turned on simultaneously when the raw water tank pump (P101) is turned on, and switched off when the raw water tank pump is turned off. In the simulation, we also require to turn on the HCl dosing pump when the measured HCl level is higher than $0.99$, and the NaOCl dosing pump is turned on when the measured NaOCl level is lower than $0.01$. In the attacking phase, we start to inject malicious HCl and NaOCl measurements at 1,000 randomly selected time points. Then, we consider an attack is successful when either the HCl or the NaOCl dosing pump is turned on unexpectedly by the injected malicious measurements when the raw water tank pump is still in the off state and meanwhile the malicious measurements must bypass the baseline anomaly detector until then (the quality of water to the third stage is changed if this happens).

\begin{figure}
  \centering
  \includegraphics[width=0.99\linewidth]{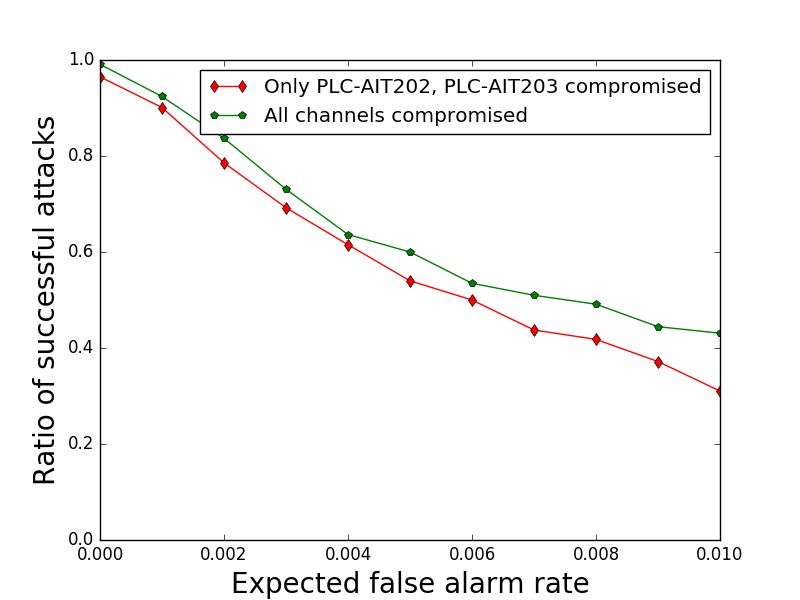}
\caption{The successful ratio of turning on the HCl or NaOCl dosing pump by the injected malicious measurements while bypassing the baseline anomaly detector with different values of expected false alarm rate}
  \label{fig:succ_residual}
\end{figure}

\begin{figure}
  \centering
  \includegraphics[width=0.99\linewidth]{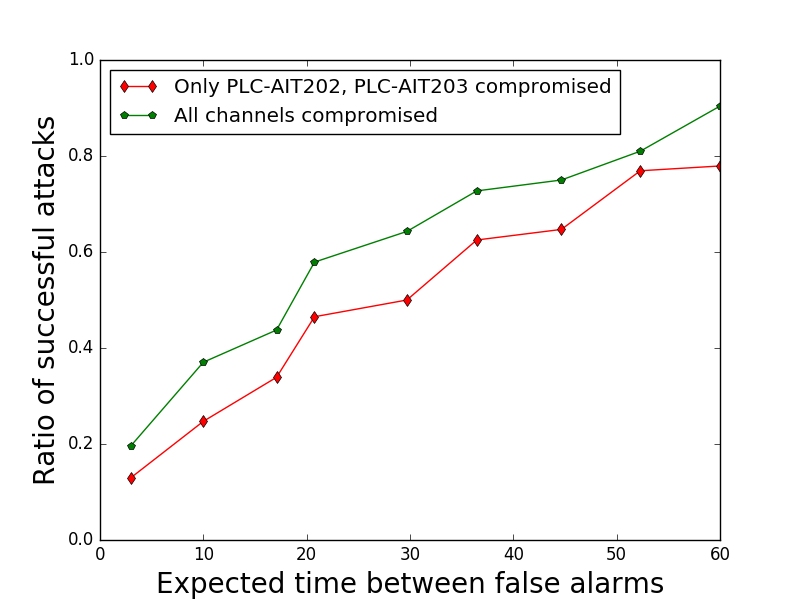}
\caption{The successful ratio of turning on the HCl or NaOCl dosing pump by the injected malicious measurements while bypassing the baseline anomaly detector with different values of expected time between false alarms (measured in minutes)}
  \label{fig:succ_cusum}
\end{figure}

The simulation results are shown in Figure~\ref{fig:succ_residual} and~\ref{fig:succ_cusum}, where we plot the ratio of successful attacks among the 1,000 attempts with different values of expected false alarm rate and expected time between false alarms, respectively. As can be seen from the figures, again, we find that only compromising the PLC-sensor channels for injecting malicious sensor measurements can still have rather high successful rates of conducting stealthy attacks. From Figure~\ref{fig:succ_residual}, we can see that the attacks will be successful on almost every attempt if the expected false alarm rate is tunned to zero (by setting the threshold $\tau$ to the maximal residual error in the training dataset). The ratios of successful attacks in both attack scenarios are still rather high (about $40\%$) even with $0.01$ expected false alarm rate (it is expected that 1 out of 100 detections will generate a false alarm). When using the CUSUM statistic to detect our attacks, we find that  the ratio of successful attacks can be as high as $90\%$ if all channels are compromised, and about $80\%$ if only the PLC-AIT202, PLC-AIT203 channels are compromised, given that the expected time between false alarms is tunned to be one hour as illustrated in Figure~\ref{fig:succ_cusum}. Even setting the expected time between false alarms to around three minutes, we can still get around $20\%$ of successful attacks which is still rather significant considering the severe consequence once the attack succeeded. From the above observations, we can clearly see that such malicious sensor measurements generated by our stealthy attack GAN can easily cause serious physical damages to ICS in the real world.


\section{Conclusion}

In this paper we have developed and demonstrated a novel GAN based stealthy attack framework, requiring a much lower \emph{a-priori} operational knowledge of a targeted ICS instance than has been previously reported for
\emph{stealthy attacks}.This feature of the framework makes it
agnostic to specific ICS types, making it applicable to a broad range
of industrial scenarios. Specifically, we have shown a real-time adversarial learning
method which provides the theoretical foundation for allowing the
attacker to inject a malicious program to automatically conduct
stealthy attacks which can bypass the deployed anomaly detector that is assumed as a black box to him. Furthermore, our framework can even allow the attacker to specify the targeted amount of deviation caused by his attacks, and let the malicious program automatically decide the optimal amount of deviation injection at each time point.

The quality of the generated malicious sensor
measurements from the deep learning method employed is assessed
quantitatively by using two real-world datasets from two industrially
distinct ICS instances. The results show that the generated malicious sensor measurements are extremely difficult to reveal by the current anomaly detection security measures. This is because our stealthy attack GAN directly learns the behaviour of the anomaly detector, thus can constantly generate malicious measurements with a relatively small residual error with respect to the value predicted by the anomaly detector. Moreover, we also find in the experiments that even with a partial knowledge of the system dynamics, our framework can still generate high-quality stealthy attacks that can potentially cause physical damage to the underlying ICS processes.

To conclude, our work indicates that with recent advances in deep
learning techniques, the widely understood reliability of existing
anomaly detection techniques is overestimated and that, by inference,
the risk of physical harm to ICS from cyber attacks is higher. In
order to counter stealthy attacks generated by our
framework, we suggest that novel security mechanisms be developed and
implemented; for example, by the use of multiple sensors for a given physical process and conduct correlation or
cross-correlation checking between multiple sensors. Such broader consideration
of methods for mitigating cyber security risks is of importance to those
operating legacy ICS instances and to the designers of next-generation
ICS systems who are involved with creating ICS architectures that are secure
by design.


\section*{Acknowledgment}
We thank iTrust, Centre for Research in Cyber Security, Singapore University of Technology and Design for providing the SWaT dataset. Cheng Feng and Deeph Chana are supported by the EPSRC project Security by Design for Interconnected Critical Infrastructures, EP/N020138/1. Tingting Li is supported by the EPSRC project RITICS: Trustworthy Industrial Control Systems, EP/L021013/1.

\bibliographystyle{IEEEtran}
\bibliography{sigproc}

\end{document}